\documentclass[aps,floats]{revtex4}
\usepackage{amsmath,amssymb}
\usepackage{graphicx,epsfig}
\usepackage[greek,english]{babel}
\usepackage{bbold}

\begin{document}
\bibliographystyle {plain}
\pdfoutput=1
\def\oppropto{\mathop{\propto}} 
\def\opsimeq{\mathop{\simeq}}
\def\opoverderline{\mathop{\overline}}
\def\operarrow{\mathop{\longrightarrow}}
\def\opsim{\mathop{\sim}}

\def\opmin{\mathop{\min}} 
\def\opmax{\mathop{\max}} 
\def\oplim{\mathop{\lim}}

\def\fig#1#2{\includegraphics[height=#1]{#2}}
\def\figx#1#2{\includegraphics[width=#1]{#2}}


\title{ Large deviations for metastable states of Markov processes with absorbing states \\
with applications to population models in stable or randomly switching environment } 


\author{ C\'ecile Monthus }
 \affiliation{Institut de Physique Th\'{e}orique, 
Universit\'e Paris Saclay, CNRS, CEA,
91191 Gif-sur-Yvette, France}

\begin{abstract}
The large deviations at Level 2.5 are applied to Markov processes with absorbing states in order to obtain the explicit extinction rate of metastable quasi-stationary states in terms of their empirical time-averaged density and of their time-averaged empirical flows over a large time-window $T$. The standard spectral problem for the slowest relaxation mode can be recovered from the full optimization of the extinction rate over all these empirical observables and the equivalence can be understood via the Doob generator of the process conditioned to survive up to time $T$. The large deviation properties of any time-additive observable of the Markov trajectory before extinction can be derived from the Level 2.5 via the decomposition of the time-additive observable in terms of the empirical density and the empirical flows. This general formalism is described for continuous-time Markov chains, with applications to population birth-death model in a stable or in a switching environment, and for diffusion processes in dimension $d$.

\end{abstract}

\maketitle


\section{ Introduction  }

Stochastic processes with absorbing states have attracted a lot of interest in various fields,
in particular to understand the properties of non-equilibrium phase transitions (see the review \cite{hinrichsen}
and references therein) and to characterize the metastable quasi-stationary states that may survive
during a very large time before extinction \cite{doorn,dickman,monasson,meleard,pollett,biblio,meleard_book,collet}.
In the field of stochastic populations models, 
the dynamical large deviations properties before extinction have been much studied 
recently via the WKB method \cite{meerson06,assaf06,meerson08,assaf08,meerson09,assaf09,assaf10a,assaf10b,meerson10,meerson12,meerson16,meerson17}.

The goal of the present paper is to offer another perspective on the 
large deviations properties of metastable quasi-stationary states :
the explicit extinction rate of metastable states will be computed
in terms of their empirical time-averaged density and of their time-averaged empirical flows over a large time-window $T$ via the applications of the large deviations at Level 2.5
to Markov processes with absorbing states.
Indeed, the initial classification of large deviations into three nested levels 
(see the reviews \cite{oono,ellis,review_touchette} and references therein),
 with Level 1 for empirical observables,
Level 2 for the empirical density,
and Level 3 for the empirical process, has been more recently supplemented
by the Level 2.5 concerning the joint distribution of the empirical density and of the empirical flows.
 The rate functions at Level 2.5 are explicit for various types of Markov processes,
including discrete-time Markov chains
 \cite{fortelle_thesis,fortelle_chain,review_touchette,c_largedevdisorder,c_reset,c_inference},
continuous-time Markov jump processes
\cite{fortelle_thesis,fortelle_jump,maes_canonical,maes_onandbeyond,wynants_thesis,chetrite_formal,BFG1,BFG2,chetrite_HDR,c_ring,c_interactions,c_open,c_detailed,barato_periodic,chetrite_periodic,c_reset,c_inference,c_runandtumble,c_jumpdiff,c_skew,c_east,c_exclusion}
and Diffusion processes 
\cite{wynants_thesis,maes_diffusion,chetrite_formal,engel,chetrite_HDR,c_reset,c_lyapunov,c_inference}.
As a consequence, the explicit Level 2.5 can be considered as a starting point
 from which many other large deviations properties can be derived
via contraction. In particular, the Level 2 for the empirical density alone 
can be obtained via the optimization of the Level 2.5 over the empirical flows,
so that the Level 2 will be closed only if this contraction can be implemented explicitly.
More generally, the Level 2.5 can be contracted
to obtain the large deviations properties of any time-additive observable
of the dynamical trajectory involving both the configuration and the flows.
The link with the studies of general time-additive observables 
via deformed Markov operators  \cite{derrida-lecture,sollich_review,lazarescu_companion,lazarescu_generic,jack_review,vivien_thesis,lecomte_chaotic,lecomte_thermo,lecomte_formalism,lecomte_glass,kristina1,kristina2,jack_ensemble,simon1,simon2,simon3,Gunter1,Gunter2,Gunter3,Gunter4,chetrite_canonical,chetrite_conditioned,chetrite_optimal,chetrite_HDR,touchette_circle,touchette_langevin,touchette_occ,touchette_occupation,garrahan_lecture,Vivo,chemical,derrida-conditioned,derrida-ring,bertin-conditioned,touchette-reflected,touchette-reflectedbis,c_lyapunov,previousquantum2.5doob,quantum2.5doob,quantum2.5dooblong,c_ruelle,lapolla}
 can be then understood via the corresponding conditioned process obtained from the generalization of Doob's h-transform.

The paper is organized as follows.
In section \ref{sec_spectral}, we introduce the notations for the
Markov jump processes with absorbing states considered in the main text,
and we recall the spectral analysis of the slowest relaxation mode before extinction,
as well as the Doob generator of the process conditioned to survive.
In section \ref{sec_2.5}, we apply the large deviations at Level 2.5 to metastable quasi-stationary states
in order to obtain their extinction rate as a function of their empirical time-averaged density 
and of their empirical time-averaged flows; 
we discuss the contractions towards the lower levels, and in particular 
we describe the link with the standard spectral analysis of section \ref{sec_spectral}.
This general framework is then illustrated with
the applications to the birth-death model in a stable environment in section \ref{sec_bd}
and in a switching environment in section \ref{sec_twolevel}.
Our conclusions are summarized in section \ref{sec_conclusion}.
The Appendix \ref{app_fp} is devoted to diffusion processes in dimension $d$ with absorbing states,
in order to explain the appropriate adaptations that are needed
with respect to the formalism explained in the main text for Markov jump processes.


\section{ Reminder on the spectral analysis of the slowest relaxation mode   }

\label{sec_spectral}

In this section, we recall the standard spectral analysis of the slowest relaxation mode 
for Markov jump processes with absorbing states \cite{doorn,dickman,monasson,meleard,pollett,biblio,meleard_book,collet}
that will be useful to understand the large deviations properties described in section \ref{sec_2.5}.

\subsection{ Markov jump process with an absorbing state at configuration $x=0$}

In the main text of this paper, we consider a Markov chain 
in continuous-time $t$ for a discrete space $\Omega$ of configurations,
with a single absorbing state that will be called the dead configuration $x=0$, 
while $\Omega^* = \{ x \in \Omega, x \ne 0\}$ represents the set of living configurations.
The generator $w$ of the Markov chain is a matrix of size $\Omega \times \Omega$,
where the off-diagonal $x \ne y$ matrix element $w_{x,y} \geq 0$   
represents the transition rate from configuration $y$ to configuration $x $,
while the diagonal element $w_{x,x}$ is negative and is fixed by the conservation of probability to be
\begin{eqnarray}
w_{x,x} \equiv  - \sum_{y \ne x} w_{y,x}
\label{wdiag}
\end{eqnarray}
The master equation for the probability $P_{t}( x )  $ to be in configuration $x$ at time $t$ reads
\begin{eqnarray}
 \partial_t P_{t}( x )    =   \sum_{y }  w_{x,y} P_t(y) = \sum_{y \ne x } \left[ w_{x,y} P_t(y) -  w_{y,x} P_t(x) \right]
\label{master}
\end{eqnarray}
The absorbing character of the dead configuration $x=0$ means that all the rates out of $x=0$ vanish
\begin{eqnarray}
 w_{y,0} =0
\label{0abs}
\end{eqnarray}
so that the master equation of Eq. \ref{master} for the dead configuration $x=0$
 reduces to the incoming flows from the living configurations $y \ne 0 $ with non-vanishing absorbing rates $w_{0,y}>0$ 
\begin{eqnarray}
 \partial_t P_{t}( 0 )    =    \sum_{y \ne 0 } w_{0,y} P_t(y) 
\label{master0}
\end{eqnarray}
As a consequence, the dynamics will converge for large time $t \to +\infty$ towards the 
dead configuration $x=0$
\begin{eqnarray}
 P_{t=+\infty}(x) = \delta_{x,0}
\label{masterabs}
\end{eqnarray}
Eqs   \ref{wdiag}  and \ref{masterabs}  mean that 
 the highest eigenvalue of the Markov matrix $w$ is zero 
 \begin{eqnarray}
 0 && = \langle l_0 \vert w = \sum_x l_0(x) w_{x,y}
 \nonumber \\
 0  && =  w \vert r_0 \rangle =  \sum_{y }  w_{x,y} r_0(y)
\label{mastereigen0}
\end{eqnarray}
where the positive left eigenvector is constant over the whole space $\Omega$ of configurations 
\begin{eqnarray}
 l_0(x)=1
\label{markovleft}
\end{eqnarray}
while the positive right eigenvector is the trivial steady state of Eq. \ref{masterabs}
\begin{eqnarray}
 r_0(y)=  \delta_{y,0}
\label{markovright}
\end{eqnarray}
with the normalization
\begin{eqnarray}
1 = \langle l_0 \vert r_0 \rangle = \sum_x l_0(x) r_0(x) = 1
\label{markovrightleftnorma}
\end{eqnarray}


\subsection{ Slowest relaxation mode with its decay rate $\zeta_1$, its right eigenvector $\vert r_1 \rangle $ and its left eigenvector $\langle l_1 \vert $ }

We are interested in models 
 where the next eigenvalue $(-\zeta_1)<0$
of the Markov matrix $w$ is very close to zero,
with its right eigenvector 
$\vert r_1 \rangle $ and its left eigenvector $ \langle l_1  \vert  $
\begin{eqnarray}
 w \vert r_1 \rangle && = -   \zeta_1 \vert r_1 \rangle 
 \nonumber \\
 \langle l_1  \vert  w  && = -   \zeta_1  \langle l_1  \vert 
\label{eigen1}
\end{eqnarray}
This slowest relaxation mode will govern the convergence towards the absorbing state of Eq. \ref{masterabs}
\begin{eqnarray}
 e^{wt }  \opsimeq_{t \to +\infty}  \vert r_0 \rangle \langle l_0  \vert  +   e^{- t \zeta_1 } \vert r_1 \rangle \langle l_1  \vert 
\label{expwspectral}
\end{eqnarray}
i.e. for the probability $ P_t(x,x_0)$ to be at configuration $x$ at time $t$ 
when starting at configuration $x_0$ at time $t=0$
\begin{eqnarray}
P_t(x,x_0) \equiv \langle x \vert e^{wt} \vert x_0 \rangle 
\opsimeq_{t \to +\infty} r_0(x) l_0(x_0) +  e^{- t \zeta_1 } r_1(x) l_1(x_0)
= \delta_{x,0}  +  e^{- t \zeta_1 } r_1(x) l_1(x_0)
\label{jumppropa}
\end{eqnarray}
Let us write more explicitly the projections of the eigenvalue Equations \ref{eigen1}
onto the dead configuration $x=0$ 
\begin{eqnarray}
-\zeta_1 r_1(0) && = \sum_y  w_{0,y}   r_1 (y) = \sum_{y \ne 0}   w_{0,y}   r_1 (y) 
 \nonumber \\
-\zeta_1 l_1(0) && = \sum_x l_1(x) w_{x,0}  =0
\label{eigen1dead}
\end{eqnarray}
and the projection onto the living configurations of $\Omega^*$ using Eq. \ref{0abs}
and \ref{eigen1dead}
\begin{eqnarray}
-\zeta_1 r_1(x) && = \sum_y  w_{x,y}   r_1 (y) = \sum_{y \ne 0}   w_{x,y}   r_1 (y) \ \ {\rm for } \ \ x \ne 0
 \nonumber \\
-\zeta_1 l_1(y) && = \sum_x l_1(x) w_{x,y}  =   \sum_{x \ne 0} l_1(x) w_{x,y}        \ \ {\rm for } \ \ y \ne 0
\label{eigen1living}
\end{eqnarray}
 as well as the orthogonality conditions with the eigenvectors of Eqs \ref{markovleft} and \ref{markovright}
\begin{eqnarray}
0 && = \langle  l_0 \vert r_1 \rangle = \sum_x l_0(x) r_1(x) =  \sum_x  r_1(x) = r_1(0) +  \sum_{x \ne 0}   r_1(x)
\nonumber \\
0 && = \langle  l_1 \vert r_0 \rangle = \sum_x l_1(x) r_0(x) = l_1(0) 
  \label{ortho01}
\end{eqnarray}
and the normalization 
\begin{eqnarray}
1 = \langle  l_1 \vert r_1  \rangle  = \sum_x l_1(x) r_1(x) = \sum_{x \ne 0}  l_1(x) r_1(x)
  \label{norma1living}
\end{eqnarray}

Let us now summarize the two important properties of the previous discussion : 

(i) the right eigenvector $r_1(.)$ and the left eigenvector $l_1(.)$ satisfy the closed eigenvalues Eqs. 
\ref{eigen1living} on the set $\Omega^* $ of living configurations with the normalization of Eq. \ref{norma1living}.
Since $(-\zeta_1)<0$ is the highest eigenvalue of Markov matrix $w$ on the set $\Omega^* $ of living configurations,
one obtains via the Perron-Frobenius that the eigenvectors $r_1$ and $l_1$ are positive on $\Omega^*$
\begin{eqnarray}
 && r_1(x)  >0 \ \ \ \ {\rm for } \ \ \ \ x \ne 0 
 \nonumber \\
  && l_1(x)  >0 \ \ \ \ {\rm for  } \ \ \ \  x \ne 0 
  \label{r1l1positive}
\end{eqnarray}

(ii) For the dead configuration $x=0$, the left eigenvector vanishes $l_1(0)=0$ (Eqs \ref{eigen1dead} and \ref{ortho01}),
while the right eigenvector $r_1(0)$ is negative and can be computed in terms of the components $r_1(x)$ on living configurations
via Eqs \ref{eigen1dead} and \ref{ortho01} 
\begin{eqnarray}
 r_1(0) && =  - \frac{1}{\zeta_1}   \sum_{y \ne 0}   w_{0,y}   r_1 (y) <0
\nonumber \\
 r_1(0) && = -  \sum_{x \ne 0}   r_1(x) <0
  \label{r10neg}
\end{eqnarray}
As a consequence, one can eliminate $r_1(0)$ 
to obtain the following consistency equation between the eigenvalue $\zeta_1 $
and the right eigenvector $r_1(x)$ for living configurations $x \ne 0$
\begin{eqnarray}
\zeta_1 = \frac{ \displaystyle  \sum_{y \ne 0}   w_{0,y}   r_1 (y)  }{ \displaystyle  \sum_{x \ne 0}   r_1(x) }
  \label{zeta1consistency}
\end{eqnarray}


\subsection{ Probability $ P^{end}_t(x,x_0) $ to be in the living configuration $x$ at time $t$ if surviving up to time $t$    }

The probability $S_t(x_0)$ to be surviving at time $t$ 
when starting at configuration $x_0$ at time $t=0$ 
can be computed from Eq. \ref{jumppropa} 
\begin{eqnarray}
S_t(x_0) && = \sum_{x \ne 0}  P_t(x,x_0) \opsimeq_{t \to +\infty}
  e^{- t \zeta_1 } \left[ \sum_{x \ne 0} r_1(x) \right] l_1(x_0)
\label{k1leading}
\end{eqnarray}
So the probability $ P^{end}_t(x,x_0) $ to be in the living configuration $x \ne 0 $ at time $t$ if surviving up to time $t$ 
\begin{eqnarray}
P^{end}_t(x,x_0) = \frac{P_t(x,x_0)  } {S_t(x_0) } \opsimeq_{t \to +\infty}  \frac{ r_1(x) } { \displaystyle \sum_{x' \ne 0} r_1(x')} \equiv 
\pi^{end}(x)
\label{endconditioned}
\end{eqnarray}
is actually independent of the time $t$ and of the initial condition $x_0$ : 
the probability distribution $\pi^{end}(x) $ normalized on the set $\Omega^*$
of living configurations thus only involves the right eigenvector $r_1(.)$ of the slowest relaxation mode.


\subsection{ Probability $ P^{interior}_{\tau}(x) $ to be in configuration $x$ at the interior time $1 \ll \tau \ll t $ if surviving up to time $t$ }

If the process survives up to time $t$, the probability $ P^{interior}_{\tau}(x) $ to be in the living configuration $x$ at the interior time $\tau$ satisfying $1 \ll \tau \ll t $  can be evaluated using the asymptotic form of the propagator of Eq. \ref{jumppropa}
for both time-intervals $\tau \gg 1 $ and $(t-\tau) \gg 1$
\begin{eqnarray}
P^{interior}_{\tau}(x) && = \frac{ P_{t-\tau} (x_t,x) P_{\tau} (x,x_0)}
{ \displaystyle \sum_{x'\ne 0} P_{t-\tau} (x_t,x') P_{\tau} (x',x_0) }
\opsimeq_{ 1 \ll \tau \ll t } 
\frac{ e^{- \zeta_1 (t-\tau) }  r_1(x_t) l_1(x)  e^{- \zeta_1 \tau}  r_1(x) l_1(x_0) }
{ \displaystyle  \sum_{x'\ne 0} e^{- \zeta_1 (t-\tau) }  r_1(x_t) l_1(x')  e^{- \zeta_1 \tau}  r_1(x') l_1(x_0)}
\nonumber \\
&& \opsimeq_{ 1 \ll \tau \ll t } 
\frac{    l_1(x)    r_1(x)  }
{ \displaystyle  \sum_{x'\ne 0}   l_1(x')    r_1(x')  } =  l_1(x)    r_1(x)  \equiv \pi^{interior}(x)
\label{interiorconditioned}
\end{eqnarray}
where we have used the normalization of Eq. \ref{norma1living} to obtain the final result.
Since the probability $P^{interior}_{\tau}(x,x_0) $ of Eq. \ref{interiorconditioned} is independent of $\tau$
 as long as it is in the interior of the time-interval $1 \ll \tau \ll t $,
it is interesting to construct the Doob Markov generator ${ \hat w }$ on the set $\Omega^*$
of living configurations that has $ \pi^{interior}(x)$ as true steady state,
i.e. its highest eigenvalue zero should be associated to the positive left and right eigenvectors
\begin{eqnarray}
{\hat l}_0 (x) && = 1
\nonumber \\
{\hat r}_0 (x) && =\pi^{interior}(x) = l_1(x) r_1(x) 
\label{rhocond}
\end{eqnarray}
The  explicit form of the Doob generator ${ \hat w } $
involves the eigenvalue $(-\zeta_1)$ and the positive left eigenvector $l_1(.)$ on $\Omega^* $
\begin{eqnarray}
{ \hat w }_{x,y} =  l_1(x) w_{x,y} \frac{1}{l_1(y) } + \zeta_1 \delta_{x,y}
\label{doob}
\end{eqnarray}
The eigenvalues equations can be checked using Eqs \ref{eigen1living}
\begin{eqnarray}
\sum_{y \ne 0} { \hat w }_{x,y} {\hat r}_0 (y)
&&  = 
\sum_{y \ne 0} \left[ l_1(x) w_{x,y} \frac{1}{l_1(y) } + \zeta_1 \delta_{x,y} \right]  l_1(y) r_1(y) 
=  l_1(x) \left[ \sum_{y \ne 0}  w_{x,y} r_1(y) \right]
+   \zeta_1    l_1(x) r_1(x) =0
\nonumber \\
\sum_{x \ne 0} {\hat l}_0 (x) { \hat w }_{x,y} 
&& =  
\sum_{x \ne 0} \left[ l_1(x) w_{x,y} \frac{1}{l_1(y) } + \zeta_1 \delta_{x,y} \right] 
= \left[ \sum_{x \ne 0}  l_1(x) w_{x,y} \right] \frac{1}{l_1(y) } + \zeta_1 =0
\label{doobcheck}
\end{eqnarray}


\subsection{ Perturbation theory with respect to the absorbing rates $w_{0,y}$ towards the dead configuration $x=0$ }

Since the spectral problem for the slowest relaxation mode cannot be explicitly solved in many models of interest,
it is useful to consider the simplest possible approximation, namely the 
perturbation theory with respect to the absorbing rates $w_{0,y}$ towards the dead configuration $x=0$.


\subsubsection{ Properties of the unperturbed Markov matrix $w^{(0)}  $   }

Let us decompose the full Markov matrix $w$ into the two contributions
\begin{eqnarray}
w = w^{(0)} + \epsilon  w^{(1)}  
\label{diadAB}
\end{eqnarray}
where the contribution containing all the absorbing rates $w_{0,y}$ towards the dead configuration $x=0$
\begin{eqnarray}
\epsilon w^{(1)}  = \sum_{ y  \ne 0  } \left( \vert 0 \rangle - \vert y \rangle \right) w_{0,y} \langle y \vert 
\label{w1per}
\end{eqnarray}
will be considered as a perturbation of order $\epsilon$
with respect to the complementary contribution containing all the other transition rates $w_{x,y} $ within the set 
$\Omega^*$ of living configurations
\begin{eqnarray}
w^{(0)} = \sum_{x \ne 0 } \sum_{y \ne 0}  \left( \vert x \rangle - \vert y \rangle \right) w_{x,y} \langle y \vert
\label{w0unperturbed}
\end{eqnarray}
This unperturbed Markov matrix $w^{(0)} $ conserve the probability on $\Omega^*$.
For the Markov matrix $w^{(0)} $ on the full configuration space $\Omega$,
the eigenvalue $0$ is thus doubly degenerate, with the two left trivial eigenvectors
\begin{eqnarray}
l_0^{dead}(x) &&=  \delta_{x,0}
\nonumber \\
l_0^{living}(x) &&=  1-\delta_{x,0} 
\label{l0AABB}
\end{eqnarray}
The corresponding right eigenvectors are
\begin{eqnarray}
r_0^{dead}(x) &&=  \delta_{x,0}
\nonumber \\
r_0^{living}(x) &&=  (1-\delta_{x,0} ) \pi^*(x)
\label{r0AABB}
\end{eqnarray}
where $ \pi^*(x)$ is the normalized steady state of Markov matrix $w^{(0)} $ on $\Omega^*$.


\subsubsection{  Degenerate perturbation theory for the unperturbed Markov matrix $w^{(0)}  $  }

The goal is to compute the leading contribution for the non-vanishing eigenvalue $(- \zeta_1 (\epsilon) )<0$ for $\epsilon>0$ of the full matrix $w$ that vanishes for $\epsilon=0$
\begin{eqnarray}
\zeta_1 (\epsilon) = 0 +\epsilon \zeta_1^{(1)} +O(\epsilon^2)
  \label{e1series}
\end{eqnarray}
with the corresponding expansions for the right and left eigenvectors
\begin{eqnarray}
\vert r_1 (\epsilon) \rangle && = \vert r_1^{(0)}  \rangle +\epsilon \vert r_1^{(1)}\rangle  +O(\epsilon^2)
\nonumber \\
\langle  l_1 (\epsilon) \vert && = \langle  l_1^{(0)}  \vert  +\epsilon \langle  l_1^{(1)}  \vert  +O(\epsilon^2)
  \label{rl1series}
\end{eqnarray}
that should satisfy 
the orthonormalization relations
\begin{eqnarray}
0 && = \langle  l_0 \vert r_1 (\epsilon) \rangle =
\langle  l_0 \vert r_1^{(0)}  \rangle +\epsilon \langle  l_0 \vert r_1^{(1)}\rangle +O(\epsilon^2)
\nonumber \\
0 && = \langle  l_1 (\epsilon) \vert r_0 \rangle =
\langle  l_1^{(0)}  \vert r_0 \rangle +\epsilon \langle  l_1^{(1)}  \vert r_0 \rangle  +O(\epsilon^2)
\nonumber \\
1&& = \langle  l_1 (\epsilon)\vert r_1 (\epsilon) \rangle  = 
\langle  l_1^{(0)}  \vert r_1^{(0)}  \rangle
+ \epsilon \left( \langle  l_1^{(0)}  \vert r_1^{(1)}  \rangle  + \langle  l_1^{(1)}  \vert r_1^{(0)}  \rangle\right) 
+O(\epsilon^2)
  \label{orthodege}
\end{eqnarray}

Since the eigenvalue $\zeta_0^{(0)}=0 $ of the unperturbed matrix $w^{(0)} $ is doubly degenerate,
one needs to apply the degenerate perturbation theory familiar from quantum mechanics,
where the first task is the diagonalization of the perturbation $(\epsilon w^{(1)} )$ within the two-dimensional degenerate subspace 
of $w^{(0)} $.
However here, we know that $\zeta_0=0$ is an exact isolated eigenvalue for the full Markov matrix $w$ for $\epsilon >0$,
where the corresponding trivial left eigenvector $l_0(x)=1$ of Eq. \ref{markovleft} is the simple linear combination of the two left trivial eigenvectors of Eq. \ref{l0AABB}
\begin{eqnarray}
\langle  l_0  \vert  &&  =  \langle  l_0^{dead} \vert +   \langle  l_0^{living} \vert
  \label{linearableft}
\end{eqnarray}
while the corresponding right eigenvector $r_0(x)=\delta_{x,0}$ of Eq. \ref{markovright}
simply coincide with one of the two right eigenvectors of Eq. \ref{r0AABB}
\begin{eqnarray}
 \vert r_0  \rangle &&  =   \vert r_0^{dead}  \rangle  
  \label{linearab}
\end{eqnarray}

At order $0$, the eigenvectors $\vert r_1^{(0)} \rangle $ and $\langle  l_1^{(0)}  \vert $ 
of Eq. \ref{rl1series} belong to the two-dimensional degenerate subspace of $w^{(0)}$ 
and can be thus rewritten as linear combinations of the corresponding unperturbed eigenvectors.
The orthonormalization conditions of Eq. \ref{orthodege} for $\epsilon=0$
\begin{eqnarray}
0 && =\langle  l_0 \vert r_1^{(0)}  \rangle  
\nonumber \\
0 &&  =\langle  l_1^{(0)}  \vert r_0 \rangle 
\nonumber \\
1&& = \langle  l_1^{(0)}  \vert r_1^{(0)}  \rangle
  \label{orthodege0}
\end{eqnarray}
 leads to the linear combinations
\begin{eqnarray}
 \vert r_1^{(0)}  \rangle &&  =  - \vert r_0^{dead}  \rangle +  \vert r_0^{living}  \rangle 
\nonumber \\
\langle  l_1^{(0)}  \vert  &&  =  \langle  l_0^{living} \vert
  \label{linear1ab}
\end{eqnarray}


\subsubsection { First-order contribution to the slowest relaxation rate $\zeta_1$  }

Using the series expansion of Eqs \ref{e1series}
and \ref{rl1series},
the eigenvalue equation for $\zeta_1 (\epsilon) $ reads for the right eigenvector $\vert r_1 (\epsilon) \rangle  $
\begin{eqnarray}
 0  &&  = \left( w + \zeta_1 (\epsilon)  \right) \vert r_1 (\epsilon) \rangle 
 = \left( w^{(0)} +\epsilon w^{(1)} + \epsilon \zeta_1^{(1)} +...  \right)  \left( \vert r_1^{(0)}  \rangle +\epsilon \vert r_1^{(1)}\rangle  +...\right)
 \label{powerright}
\\
 && =  \epsilon
\left[ w^{(0)}  \vert r_1^{(1)} \rangle + \left( w^{(1)} + \zeta_1^{(1)}  \right) \vert r_1^{(0)} \rangle \right]
+O(\epsilon^2)
 \nonumber 
\end{eqnarray}
and for the left eigenvector $  \langle  l_1 (\epsilon)$
\begin{eqnarray}
0  && = \langle  l_1 (\epsilon)\vert\left( w + \zeta_1 (\epsilon)  \right) 
= \left( 
\langle  l_1^{(0)}  \vert  +\epsilon \langle  l_1^{(1)}  \vert +...  
  \right)
 \left( w^{(0)} +\epsilon w^{(1)} + \epsilon \zeta_1^{(1)}+ ...  \right)
  \label{powerleft}
 \\
&& = 
 \epsilon
\left[ \langle  l_1^{(1)} \vert w^{(0)}   +\langle  l_1^{(0)} \vert  \left( w^{(1)} + \zeta_1^{(1)}  \right) 
\right]
+O(\epsilon^2)
  \nonumber
\end{eqnarray}
The standard choice that respects the normalization of Eq \ref{orthodege} at order $\epsilon$
is given by the orthogonality conditions for the first-order corrections
of the eigenvectors with respect to the unperturbed eigenvectors
\begin{eqnarray}
0 && =  \langle  l_1^{(0)}  \vert r_1^{(1)}  \rangle  
\nonumber \\
0 && =  \langle  l_1^{(1)}  \vert r_1^{(0)}  \rangle
  \label{powernorma1}
\end{eqnarray}
Then the eigenvalue Eq. \ref{powerright} for the right eigenvector at order $\epsilon$
can be projected onto the left eigenvector $\langle  l_1^{(0)} \vert $ satisfying $\langle  l_1^{(0)} \vert w^{(0)} =0 $ 
and $\langle  l_1^{(0)} \vert r_1^{(0)} \rangle =1 $
to obtain the first-order correction for the eigenvalue
\begin{eqnarray}
 \zeta_1^{(1)} = - \langle  l_1^{(0)} \vert w^{(1)}  \vert r_1^{(0)} \rangle
  \label{energy1}
\end{eqnarray}
Equivalently, the eigenvalue Eq. \ref{powerleft} for the leftt eigenvector at order $\epsilon$
can be projected onto the right eigenvector $\vert r_1^{(0)} \rangle$ 
satisfyinf $w^{(0)}  \vert r_1^{(0)} \rangle =0 $ 
to obtain again Eq. \ref{energy1}.

Putting everything together, this degenerate perturbation theory yields that the leading contribution 
for the slowest relaxation rate can be evaluated from the matrix elements of the perturbation $\epsilon w^{(1)}
 $ of Eq. \ref{w1per}
for the left and right eigenvectors of Eq. \ref{linear1ab}
\begin{eqnarray}
\zeta_1^{per} &&  = \epsilon \zeta_1^{(1)} = - \langle  l_1^{(0)} \vert \epsilon w^{(1)}  \vert r_1^{(0)} \rangle
 =  \sum_{ y  \ne 0  }  \langle  l_0^{living} \vert \left( - \vert 0 \rangle + \vert y \rangle \right) w_{0,y} \langle y \vert \left(   - \vert r_0^{dead}  \rangle +  \vert r_0^{living}  \rangle\right)
 \nonumber \\ &&
=   \sum_{ y  \ne 0  }   l_0^{living} (y) w_{0,y}  r_0^{living}(y) 
=   \sum_{ y  \ne 0  } w_{0,y}  \pi^*(y) 
  \label{zeta1per}
\end{eqnarray}
This final result has thus a very simple physical meaning in terms of the absorbing flows $w_{0,y}  \pi^*(y) $
into the dead configuration $x=0$ computed with the normalized steady-state
$\pi^*(y) $ of the unperturbed Markov matrix $w^{(0)} $ of Eq. \ref{w0unperturbed}
on the set $\Omega^*$ of living configurations.
As a consequence, even when the absorbing rates $w_{0,y} $ are not particularly small,
the perturbative estimation $\zeta_1^{per}  $ of Eq. \ref{zeta1per}
will be small if the steady state $ \pi^*(y)$ is small on the sites $y \ne 0$ that are directly connected to
the absorbing site at $x=0$ via positive rates $w_{0,y}>0$.


\section{ Large deviations at various levels for metastable quasi-stationary states   }

\label{sec_2.5}

In this section, the goal is to analyze the extinction rate of 
metastable states in terms of their empirical time-averaged properties over a large time-window $T$.


\subsection{ Reminder on the large deviations at Level 2.5 for the time-averaged density and the time-averaged flows  }

For a very long dynamical trajectory $x(0 \leq t \leq T)$
of the Markov jump process satisfying a master equation of the form of Eq. \ref{master},
the empirical time-averaged density  
\begin{eqnarray}
 \rho_T( x ) \equiv \frac{1}{T} \int_0^T dt \ \delta_{x(t) , x }
\label{rhoc}
\end{eqnarray}
measures the histogram of the various configurations $x$
seen during the dynamical trajectory $x(0 \leq t \leq T) $,
while the empirical time-averaged flows 
\begin{eqnarray}
q_T(x,y) \equiv  \frac{1}{T} \sum_{t\in [0,T] : x(t^+) \ne x(t^-)} \delta_{x(t^+),x} \delta_{x(t^-),y} 
\label{jumpempiricaldensity}
\end{eqnarray}
measure the density of jumps from one configuration $y$ to another configuration $x$.
The joint probability to see the empirical density
 $\rho( x )$ and the empirical flows $q(x,y) $
follows the
large deviation form for large $T$ 
\cite{fortelle_thesis,fortelle_jump,maes_canonical,maes_onandbeyond,wynants_thesis,chetrite_formal,BFG1,BFG2,chetrite_HDR,c_ring,c_interactions,c_open,c_detailed,barato_periodic,chetrite_periodic,c_reset,c_inference,c_runandtumble,c_jumpdiff,c_skew,c_east,c_exclusion}
\begin{eqnarray}
P^{[2.5]}_{T}[ \rho(.) ; q(.,.) ] \oppropto_{T \to +\infty} C_{2.5}[ \rho(.) ; q(.,.) ]  e^{- T I_{2.5}[ \rho(.) ; q(.,.) ] }
\label{level2.5}
\end{eqnarray}
where the prefactor
\begin{eqnarray}
C_{2.5}[ \rho(.) ; q(.,.) ] = \delta \left( \sum_x \rho(x) - 1 \right) 
 \prod_x \delta \left(  \sum_{y \ne x} \left( q(x,y) - q(y,x) \right) \right)
\label{c2.5}
\end{eqnarray}
contains the constitutive constraints that the empirical observables should satisfy :
 the density $\rho(.)$ should be normalized,
 while the flows $ q(.,.)$ should satisfy the stationarity constraints :
 for any configuration $x$, the total incoming flow $ \sum_{y \ne x}q(x,y)$ into the configuration $x$ 
 should be equal to the total outgoing flow $\sum_{y \ne x}  q(y,x) $ out of the configuration $x$.
As stressed in the Introduction, the rate function $I_{2.5}[ \rho(.) ; q(.,.) ]  $ at level 2.5 
can be written explicitly for any continuous-time Markov chain in terms of its generator $w$
\cite{fortelle_thesis,fortelle_jump,maes_canonical,maes_onandbeyond,wynants_thesis,chetrite_formal,BFG1,BFG2,chetrite_HDR,c_ring,c_interactions,c_open,c_detailed,barato_periodic,chetrite_periodic,c_reset,c_inference,c_runandtumble,c_jumpdiff,c_skew,c_east,c_exclusion}
\begin{eqnarray}
I_{2.5}[ \rho(.) ; q(.) ]=  \sum_{x } \sum_{y \ne x} 
\left[ q(x,y)  \ln \left( \frac{ q(x,y)  }{  w_{x,y}  \rho(y) }  \right) 
 - q(x,y)  + w_{x,y}  \rho(y)  \right]
\label{rate2.5}
\end{eqnarray}
This rate function characterizes how rare it is for large $T$
to see some empirical density $\rho(.)$ and some empirical flows $q(.,.)$ 
that are different from the steady state and its corresponding steady flows that would make the rate function vanish.
As a consequence, the large deviations at Level 2.5 can be 
directly applied to Markov jump processes with absorbing states
in order to analyze the extinction rate of metastable states as explained below. 


\subsection{ Probability $P^{[2.5]living}_T[ \rho(.) ; q(.,.) ]  $ to remain living up to $T$ with given empirical density and empirical flows  }

The probability to remain confined in the set $\Omega^*$ 
of living configurations during the whole time-window $T$
can be directly obtained from the Level 2.5 of Eq. \ref{level2.5}
\begin{eqnarray}
P^{[2.5]living}_T[ \rho(.) ; q(.,.) ] 
\oppropto_{T \to +\infty} C^{living}_{2.5}[ \rho(.) ; q(.,.) ]  e^{- T I^{living}_{2.5}[ \rho(.) ; q(.,.) ] }
\label{level2.5b}
\end{eqnarray}
as a function of the empirical density $\rho(.) $ and the empirical flows $q(.,.) $ defined on $\Omega^*$,
i.e. the empirical density should vanish on the dead configuration $x=0$, 
and all the empirical absorbing flows $q(0,y)$ should also vanish
\begin{eqnarray}
\rho(0) && =0
\nonumber \\
q(0,y) && =0 \ \ {\rm for \ any } \ \ y \ne 0
\label{empidead}
\end{eqnarray}
So the constraints obtained from Eq \ref{c2.5} become
\begin{eqnarray}
C^{living}_{2.5}[ \rho(.) ; q(.,.) ] 
 = \delta \left(  \rho(0) \right) \delta \left(  \sum_{x \ne 0 } \rho(x)- 1 \right) 
 \left[ \prod_{y \ne 0} \delta \left( q(0,y)  \right)\right]
\left[ \prod_{x \ne 0} \delta \left(   \sum_{ \substack{ y \ne 0 \\ y \ne x }} \left( q(x,y) - q(y,x) \right)  \right) \right]
\label{c2.5b}
\end{eqnarray}
while the rate function obtained from Eq. \ref{rate2.5} reads
\begin{eqnarray}
 I^{living}_{2.5}[ \rho(.) ; q(.,.) ] 
 =     \sum_{y \ne 0 }  w_{0,y}  \rho(y)  
 +  \sum_{x \ne 0} \sum_{ \substack{ y \ne 0 \\ y \ne x }} 
\left[ q(x,y)  \ln \left( \frac{ q(x,y)  }{  w_{x,y}  \rho(y) }  \right)  - q(x,y)  + w_{x,y}  \rho(y)  \right]
\label{rate2.5b}
\end{eqnarray}
In conclusion, the rate function $ I^{living}_{2.5}[ \rho(.) ; q(.,.) ]  $ 
 represents the extinction rate of a metastable state
  localized on the set $\Omega^*$ of living configurations
 as a function of its empirical density $\rho(.)$ and of its empirical flows $q(.,.) $.

 Using Eq. \ref{wdiag} to rewrite for any $y \ne 0$
 \begin{eqnarray}
         w_{0,y} + \sum_{ \substack{ x \ne 0 \\ x \ne y }}   w_{x,y}  = \sum_{   x \ne y }   w_{x,y} = - w_{y,y}
\label{sum}
\end{eqnarray}
one obtains the alternative expression for the Level 2.5 rate function of Eq. \ref{rate2.5b}
 \begin{eqnarray}
 I^{living}_{2.5}[ \rho(.) ; q(.,.) ] 
 =   - \sum_{y \ne 0 } w_{y,y} \rho(y)  +  \sum_{x \ne 0} \sum_{ \substack{ y \ne 0 \\ y \ne x }} 
\left[ q(x,y)  \ln \left( \frac{ q(x,y)  }{  w_{x,y}  \rho(y) }  \right)  - q(x,y)    \right]
\label{rate2.5bbis}
\end{eqnarray}


\subsection{ Probability $P^{[2]living}_T[ \rho(.)  ]  $ to remain living up to $T$ with given empirical density   }
 
  If one wishes to compute the extinction rate of a metastable state as a function of its empirical density $\rho(.)$ alone,
  one needs to integrate the joint probability of Eq. \ref{level2.5b}
   over all the possible empirical flows $q(.,.)  $
\begin{eqnarray}
&& P^{[2]living}_T[ \rho(.)  ]  = \int {\cal D} q(.,.) P^{[2.5]living}_T[ \rho(.) ; q(.,.) ] 
\nonumber \\
&& \oppropto_{T \to +\infty} 
 \delta \left(  \rho(0) \right)
\delta \left(  \sum_{x \ne 0 } \rho(x)- 1 \right)  
\int {\cal D} q(.,.) 
 \left[ \prod_{y \ne 0} \delta \left( q(0,y)  \right)\right]
\left[  \prod_{x \ne 0} \delta \left(   \sum_{ \substack{ y \ne 0 \\ y \ne x }} \left( q(x,y) - q(y,x) \right)  \right) \right]
 e^{- T  I^{living}_{2.5}[ \rho(.) ; q(.,.) ] }
 \nonumber \\
 && \oppropto_{T \to +\infty}  \delta \left(  \rho(0) \right) \delta \left(  \sum_{x \ne 0 } \rho(x)- 1 \right) 
 e^{- T I^{living}_{2}[ \rho(.)  ] }
\label{level2b}
\end{eqnarray}
So the extinction rate $ I^{living}_{2}[ \rho(.)  ] $ at Level 2 as a function of the empirical density $\rho(.)$ alone
corresponds to the optimization of the extinction rate $  I^{living}_{2.5}[ \rho(.) ; q(.,.) ] $ at Level 2.5 
over the empirical flows $q(.,.) $ satisfying the stationarity constraints.


\subsection{ Link with the slowest relaxation mode via the probability $P^{living}_T \propto e^{- T \zeta_1} $ 
to remain living up to $T$     }

The probability $P^{living}_T  $ to remain living up to $T$ that involves the slowest decay rate $\zeta_1$
discussed in section \ref{sec_spectral}
\begin{eqnarray}
P^{living}_T   \oppropto_{T \to +\infty} e^{- T \zeta_1 }
\label{level1zeta1}
\end{eqnarray}
can also be computed via the integration the Level 2 of Eq. \ref{level2b}
over the empirical density $\rho(.)$
\begin{eqnarray}
P^{living}_T = \int {\cal D} \rho(.) P^{[2]living}_T[ \rho(.)  ]  \oppropto_{T \to +\infty} 
\int {\cal D} \rho(.)  \delta \left(  \rho(0) \right) \delta \left(  \sum_{x \ne 0 } \rho(x)- 1 \right) 
 e^{- T I^{living}_{2}[ \rho(.)  ] }  
\label{level1bfrom2}
\end{eqnarray}
By consistency, the slowest decay rate $\zeta_1$
should thus correspond to the optimization of the extinction rate $I^{living}_{2}[ \rho(.)  ] $ at Level 2
over the empirical density $\rho(.)$ satisfying the normalization constraint
\begin{eqnarray}
 \zeta_1 = \opmin_{  \displaystyle \rho(.) :  \sum_{x \ne 0 } \rho(x)= 1  }  \left(  I^{living}_{2}[ \rho(.)  ] \right)
\label{zeta1asmin}
\end{eqnarray}

When the extinction rate $I^{living}_{2}[ \rho(.)  ] $ at Level 2 cannot be explicitly obtained via the contraction of Eq. \ref{level2b},
one can always return to the explicit Level 2.5 of Eq. \ref{level2.5b}
that should be integrated over both the empirical density and the empirical flows to obtain
\begin{eqnarray}
&& P^{living}_T  = \int {\cal D} \rho(.)  \int {\cal D} q(.,.) P^{[2.5]living}_T[ \rho(.) ; q(.,.) ] 
\nonumber \\
&&  \oppropto_{T \to +\infty} 
\int {\cal D} \rho(.) 
 \delta \left(  \rho(0) \right)
\delta \left(  \sum_{x \ne 0 } \rho(x)- 1 \right) 
\int {\cal D} q(.,.) 
 \left[ \prod_{y \ne 0} \delta \left( q(0,y)  \right)\right]
\left[ 
 \prod_{x \ne 0} \delta \left(   \sum_{ \substack{ y \ne 0 \\ y \ne x }} \left( q(x,y) - q(y,x) \right)  \right)
 \right]
\nonumber \\
&&  \ \ \ \ \ \ \ \ e^{- T  I^{living}_{2.5}[ \rho(.) ; q(.,.) ] }
\label{level1bfrom2.5}
\end{eqnarray}
So the slowest decay rate $\zeta_1$
corresponds to the optimization of the explicit extinction rate $ I^{living}_{2.5}[ \rho(.) ; q(.,.) ]$ at Level 2.5
over the empirical density $\rho(.)$ satisfying the normalization constraint
and over the empirical flows $q(.,.)$ satisfying the stationarity constraints.
Let us now explain in more details the links with properties discussed in section \ref{sec_spectral}.


\subsubsection{ Link with the estimation of $\zeta_1^{per}$ obtained via the perturbation theory in the absorbing rates $w_{0,y}$}

\label{subsec_zeta1perfromlargedev}

The rate function $I^{living}_{2.5}[ \rho(.) ; q(.,.) ]  $ of Eq. \ref{rate2.5b}
allows to recover directly the perturbative estimation of the slowest relaxation rate $\zeta_1^{per}$ of Eq. \ref{zeta1per}
as follows :
when the empirical density  $\rho(y)$ coincides with the steady state $ \pi^*(y)$ of 
the unperturbed Markov matrix $w^{(0)}$ on $\Omega^*$,
and when the empirical flows $q(x,y) $ coincide with the associated steady flows $w_{x,y}  \pi^*(y) $  on $\Omega^*$,
then all the constraints of Eq. \ref{c2.5b} are satisfied, while 
the second term of the rate function of Eq. \ref{rate2.5b} vanishes,
so that Eq. \ref{level2.5b} reduces to
\begin{eqnarray}
P^{living}_T[ \rho(y)= \pi^*(y) ; q(x,y)=w_{x,y}  \pi^*(y) ] 
\oppropto_{T \to +\infty}   e^{ \displaystyle - T   \sum_{y \ne 0 }  w_{0,y}  \pi^*(y)     } =  e^{ \displaystyle - T  \zeta_1^{per}    } 
\label{level2.5bstar}
\end{eqnarray}
and one recovers directly the perturbative estimation $\zeta_1^{per}$ of Eq. \ref{zeta1per}.


\subsubsection{ Equivalence between the exact optimization of Eq. \ref{level1bfrom2.5} 
and the spectral problem of section \ref{sec_spectral}}

\label{subsec_equivalence}

Let us now consider the exact optimization problem of Eq. \ref{level1bfrom2.5}.
In the integral,
it is convenient to make the change of variables from the empirical flows $q(x,y) $
to the empirical transition rates $W(x,y)$ on $\Omega^*$
\begin{eqnarray}
 W(x,y) \equiv \frac{q(x,y)}{\rho(y)} \ \ {\rm for } \ \ x \ne y
\label{wempirical}
\end{eqnarray}
The integral of Eq. \ref{level1bfrom2.5} can be then rewritten 
using the rate function of Eq. \ref{rate2.5bbis} as
\begin{eqnarray}
&& P^{living}_T 
  \oppropto_{T \to +\infty} 
\int {\cal D} \rho(.)  \delta \left(  \rho(0) \right) \delta \left(  \sum_{y \ne 0 } \rho(y)- 1 \right) 
\int {\cal D} W(.,.) 
 \left[ \prod_{y \ne 0} \delta \left( W(0,y)  \right)\right]
 \nonumber \\
&&
\left[ 
 \prod_{x \ne 0} \delta \left(  \sum_{ \substack{ y \ne 0 \\ y \ne x }} \left( W(x,y)\rho(y)  - W(y,x)\rho(x) \right)     \right)
 \right]
 e^{\displaystyle - T    \sum_{y \ne 0 }    \rho(y)  
\left( -w_{y,y}  +   \sum_{ \substack{ x \ne 0 \\ x \ne y }} 
\left[ W(x,y)   \ln \left( \frac{ W(x,y)   }{  w_{x,y}   }  \right)  - W(x,y)    \right] \right)} \ \ 
\label{level1bwempi}
\end{eqnarray}
In order to optimize the functional in the exponential of Eq. \ref{level1bwempi}
in the presence of the constraints,
it is convenient to introduce the following Lagrangian
with the Lagrange multipliers 
$(\omega,\nu(.))$
\begin{eqnarray}
 {\cal L}  ( \rho(.) ; W(.,.)) 
  && = 
    \sum_{y \ne 0 }    \rho(y)  
\left( -w_{y,y}  +     \sum_{ \substack{ x \ne 0 \\ x \ne y }} 
\left[ W(x,y)   \ln \left( \frac{ W(x,y)   }{  w_{x,y}   }  \right)  - W(x,y)    \right] \right)
 \nonumber \\ &&
- \omega \left( \sum_{y \ne 0 } \rho(y)- 1 \right) 
- \sum_{x \ne 0}  \nu(x) \left(  \sum_{ \substack{ y \ne 0 \\ y \ne x }} \left( W(x,y)\rho(y)  - W(y,x)\rho(x) \right)     \right)
\label{lagrangiandef}
\end{eqnarray}
that can be rewritten more compactly as
\begin{eqnarray}
 {\cal L}  ( \rho(.) ; W(.,.)) 
  = \omega +    \sum_{y \ne 0 }    \rho(y)  
\left( - \omega -w_{y,y} +   \sum_{ \substack{ x \ne 0 \\ x \ne y }} 
\left[ W(x,y)   \ln \left( \frac{ W(x,y)   }{  w_{x,y}   e^{ \nu(x)  -  \nu(y) } }  \right)  - W(x,y)    \right] \right)
\label{lagrangian}
\end{eqnarray}
The optimization of this Lagrangian with respect to empirical transition rate $W(x,y)$ on $\Omega^*$ for $x \ne y$
\begin{eqnarray}
0  = \frac{ \partial  {\cal L}  ( \rho(.) ; W(.,.)) }{ \partial W(x,y)} 
 =   \rho(y)   \ln \left( \frac{ W(x,y)   }{  w_{x,y} e^{ \nu(x)  -  \nu(y) }  }  \right) 
\label{lagrangianderiw}
\end{eqnarray}
yields the optimal values
\begin{eqnarray}
 W^{opt}(x,y)  = e^{ \nu(x)  } w_{x,y}   e^{-  \nu(y) }
\label{wopt}
\end{eqnarray}
The optimization of the Lagrangian of Eq. \ref{lagrangian}
with respect to the empirical density $   \rho(y)  $
yields with the optimal transition rates obtained in Eq. \ref{wopt}
\begin{eqnarray}
0  = \frac{ \partial  {\cal L}  ( \rho(.) ; W(.,.)) }{ \partial \rho(y)} 
&& = 
- \omega  -w_{y,y} +   \sum_{ \substack{ x \ne 0 \\ x \ne y }} 
\left[ W^{opt}(x,y)   \ln \left( \frac{ W^{opt}(x,y)   }{  w_{x,y}   e^{ \nu(x)  -  \nu(y) } }  \right)  - W^{opt}(x,y)     \right]
\nonumber \\
&& = - \omega  -w_{y,y} -  \sum_{ \substack{ x \ne 0 \\ x \ne y }}  e^{ \nu(x)  } w_{x,y}   e^{-  \nu(y) }   
= - \omega   -  \sum_{  x \ne 0 }  e^{ \nu(x)  } w_{x,y}   e^{-  \nu(y) }  
\label{lagrangianderirho}
\end{eqnarray}
The optimal value of the Lagrangian of Eq. \ref{lagrangian} that determines the slowest relaxation rate $ \zeta_1$ of Eq. \ref{level1bfrom2}
reduces to the Lagrange multiplier $\omega$ 
\begin{eqnarray}
\zeta_1 = {\cal L}^{opt} = \omega 
\label{lagrangianopt}
\end{eqnarray}
The optimization Eq. \ref{lagrangianderirho} can be rewritten as the eigenvalue equation
\begin{eqnarray}
- \zeta_1 e^{\nu(y)}  =  \sum_{  x \ne 0 }  e^{ \nu(x)  } w_{x,y}  
\label{eigennu}
\end{eqnarray}
so that $ e^{ \nu(x)  }$ should correspond to the positive left eigenvector $l_1(.)$
of the matrix $w$ on $\Omega^*$ associated to the eigenvalue $(-\zeta_1)$
\begin{eqnarray}
  e^{ \nu(x)  } = l_1(x)
\label{leftnu}
\end{eqnarray}
The optimal transition rates of Eq. \ref{wopt} should satisfy the stationarity constraint for any $x \ne 0$
\begin{eqnarray}
0 && =  \sum_{ \substack{ y \ne 0 \\ y \ne x }} \left( W^{opt}(x,y)\rho(y)  - W^{opt}(y,x)\rho(x) \right)   
\nonumber \\
&& =  \left[ \sum_{  y \ne 0 } e^{ \nu(x)  } w_{x,y}   e^{-  \nu(y) }\rho(y) 
- w_{x,x} \rho(x) 
\right]
 - \left[ \sum_{  y \ne 0 }  e^{ \nu(y)  } w_{y,x}  e^{-  \nu(x) } \rho(x)
 -  w_{x,x}  \rho(x)
 \right]    
 \nonumber \\
 && 
=  e^{ \nu(x)  } \sum_{  y \ne 0 } w_{x,y}   e^{-  \nu(y) }\rho(y)
-   \left[ \sum_{  y \ne 0 }  e^{ \nu(y)  } w_{y,x}  \right]   e^{-  \nu(x) } \rho(x) 
\label{woptconstraint}
\end{eqnarray}
that can be rewritten using Eq. \ref{eigennu} for the last parenthesis
as the eigenvalue equation
\begin{eqnarray}
- \zeta_1 e^{ - \nu(x)  } \rho(x)      =   \sum_{  y \ne 0 }  w_{x,y}   e^{-  \nu(y) }\rho(y) 
\label{woptconstrainteigen}
\end{eqnarray}
So $e^{-  \nu(x) }\rho(x) $ should be the positive right eigenvector $r_1(.)$ of the matrix $w$ on $\Omega^*$ associated to the eigenvalue $(-\zeta_1)$
\begin{eqnarray}
r_1(x) =  e^{-  \nu(x) }\rho(x)  = \frac{ \rho(x) }{ l_1(x) }
\label{rightnu}
\end{eqnarray}
As a consequence, the optimal empirical density $\rho^{opt}(x)$ coincides with the interior probability distribution $\pi^{interior}(x)$ of Eq. \ref{interiorconditioned}
\begin{eqnarray}
\rho^{opt}(x) =   l_1(x) r_1(x) = \pi^{interior}(x)
\label{rhoopti}
\end{eqnarray}
Finally, the optimal empirical transition rates of Eq. \ref{wopt}
 on $\Omega^*$ for $x \ne y$ involves the left eigenvector of Eq. \ref{leftnu}
\begin{eqnarray}
 W^{opt}(x,y)  = e^{ \nu(x)  } w_{x,y}   e^{-  \nu(y) } = l_1(x)  w_{x,y} \frac{1}{l_1(y)} = {\hat w}_{x,y} \ \ {\rm for } \ \ x \ne y
\label{woptdoob}
\end{eqnarray}
and corresponds to the off-diagonal matrix elements of the Doob generator of Eq. \ref{doob}.

In conclusion, the exact optimization
 of the explicit rate function $I^{living}_{2.5}[ \rho(.) ; q(.,.) ] $ at Level 2.5
over the empirical density and the empirical flows with their constraints
is equivalent to the spectral analysis described in the previous section \ref{sec_spectral},
as it should for consistency.


\subsection{ Large deviations for general time-additive observables of the trajectory $x(0 \leq t \leq T)$ before extinction  }

\label{subsec_additive}

As mentioned in the Introduction, the large deviations at Level 2.5 
allows to derive the large deviations of any time-additive observable of the Markov trajectory
using their decomposition in terms of empirical observables.
Indeed, the empirical density $\rho(.)$ allows to 
reconstruct any time-additive observable ${\cal A}_T$ that involves some function $\alpha_x$ of the configuration $x$
\begin{eqnarray}
 {\cal A}_T   \equiv \frac{1}{T} \int_0^T dt \ \alpha_{x(t)}  = \sum_x \alpha_x \rho_x
\label{additiveAjump}
\end{eqnarray}
while the empirical flows $q(.,.)$
allows to reconstruct any time-additive observable ${\cal B}_T$ that involves some function $\beta_{x,y}$ 
\begin{eqnarray}
 {\cal B}_T  \equiv  \frac{1}{T} \sum_{t: x(t^-) \ne x(t^+) } \beta_{x(t^+),x(t^-)}  
  =  \sum_y \sum_{ x \ne y } \beta_{x,y} q_{x,y} 
\label{additiveBjumpq}
\end{eqnarray}
As a consequence, the large deviations at Level 2.5 for $P^{[2.5]living}_T[ \rho(.) ; q(.,.) ]  $  of Eq. \ref{level2.5b}
allows to analyze the large deviations of any time-additive observable of the form $({\cal A}_T+{\cal B}_T) $
for Markov trajectories living up to time $T$.


\subsection{ Generalization : large deviations at Level 2.5 for a set $A$ of absorbing configurations $a \in A$  }

Up to now, we have only considered the case of a single absorbing configuration $x=0$ for simplicity,
but it is clear that the large deviations at Level 2.5 for $P^{[2.5]living}_T[ \rho(.) ; q(.,.) ]  $ of Eq. \ref{level2.5b}
can be directly generalized to the case of a set $A$ of absorbing configurations $a \in A$ as follows.
The constraints of Eq. \ref{c2.5b} become
\begin{eqnarray}
C^{living}_{2.5}[ \rho(.) ; q(.,.) ] 
 =  \left[ \prod_{a \in A} \delta \left( \rho(a)  \right)\right]
 \delta \left(  \sum_{x \notin A } \rho(x)- 1 \right) 
  \left[ \prod_{a \in A} \prod_{y \notin A} \delta \left( q(a,y)  \right)\right]
\left[ \prod_{x \notin A } \delta \left(   \sum_{ \substack{ y \notin A  \\ y \ne x }} \left( q(x,y) - q(y,x) \right)  \right) \right]
\label{c2.5bseveral}
\end{eqnarray}
while the rate function obtained from Eq. \ref{rate2.5b} is given by
\begin{eqnarray}
 I^{living}_{2.5}[ \rho(.) ; q(.,.) ] 
 =   \sum_{a \in A}  \sum_{y \notin A }  w_{a,y}  \rho(y)  
 +  \sum_{x \notin A} \sum_{ \substack{ y \notin A \\ y \ne x }} 
\left[ q(x,y)  \ln \left( \frac{ q(x,y)  }{  w_{x,y}  \rho(y) }  \right)  - q(x,y)  + w_{x,y}  \rho(y)  \right]
\label{rate2.5bseveral}
\end{eqnarray}


\subsection{ Application to other types of metastable states  }

In this paper, we have chosen to focus on Markov processes with absorbing states.
However, the case of long-lived metastable states 
before the relaxation towards the Boltzmann equilibrium or some other steady state
has also attracted a lot of interest for many classical stochastic dynamics
\cite{gaveau1,gaveau2,gaveau3,bovier,c_metastable,c_eigenvalue,c_RGdyn,c_Dyson,c_largestbarrier,c_LRSG}
(see also \cite{metastableQuantum_short,metastableQuantum_long} for 
the notion of metastability in open quantum systems).
As a consequence, it is interesting to mention here that 
the large deviations at Level 2.5 for $P^{[2.5]living}_T[ \rho(.) ; q(.,.) ]  $ of Eq. \ref{level2.5b}
with the constraints of Eq. \ref{c2.5bseveral}
and the rate function of Eq. \ref{rate2.5bseveral}
can be also directly applied to other types of metastability : the set $A$ of absorbing states 
should be then chosen as the outer boundary of the domain where the metastable state of interest lives.


\section{  Large deviations at various levels for population birth-death models  }

\label{sec_bd}

In this section, the goal is to apply the general formalism described in the two previous sections
to the birth-death model which is one of the simplest stochastic model for the continuous-time dynamics
of a discrete population $n$.

\subsection{ Birth-death model for the discrete population $n$ in the configuration space $\Omega=\{0,1,2,.,N\}$ }

The Markov generator is parametrized by the birth rates $b_m$ and the death rates $d_m$
\begin{eqnarray}
  w_{n,m} =  \left[ \delta_{n,m+1}-\delta_{n,m}\right] b_{m}
  +   \left[ \delta_{n,m-1}-\delta_{n,m}\right] d_m
\label{matrixbd}
\end{eqnarray}
with the following boundary conditions :

(i) the dead state $n=0$ is an absorbing state from which there is no escape
\begin{eqnarray}
b_0=0=d_0
\label{bd0}
\end{eqnarray}

(ii) the maximal state $n=N$ is characterized by the vanishing birth rate
\begin{eqnarray}
b_N=0
\label{bdN}
\end{eqnarray}
so that it is not possible to reach the state $n=N+1$.
One may also consider that there is no maximal population $N \to +\infty$.

So the master Eq. \ref{master} reads
\begin{eqnarray}
\partial_t  P_t(n) =   \sum_m w_{n,m} P_t(m) 
 =     b_{n-1} P_t(n-1)   +  d_{n+1} P_t(n+1)   -  (b_{n} +d_n) P_t(n)  
\label{masterbd}
\end{eqnarray}
For the dead configuration $n=0$ with the vanishing rates of Eq. \ref{bd0}, 
the dynamics reduces to the incoming flow from the neighboring site $n=1$
\begin{eqnarray}
\partial_t  P_t(0)  =         d_{1} P_t(1)   
\label{masterbdzero}
\end{eqnarray}

In various applications, many different choices for the dependence in $n$ of the birth rates $b_n$
 and of the death rates $d_n$ are relevant. As a consequence, it is interesting to discuss the general case with arbitrary birth and death rates.


\subsection{ Spectral analysis of the slowest relaxation mode of section \ref{sec_spectral}  }

\label{subsec_spectralbd}

For the present birth-death model, the eigenvalues equations of Eqs \ref{eigen1living}
on $\Omega^*=\{1,2,..,N\}$
for the slowest relaxation mode  
read for the right eigenvector $r(.)$ in the bulk $n=2,..,N-1$
\begin{eqnarray}
   - \zeta_1   r_1(n)  =   b_{n-1} r_1(n-1)   +  d_{n+1} r_1(n+1)   -  (b_{n} +d_n) r_1(n)  
      \label{rightbd}
\end{eqnarray}
with the boundary equations for $n=1$ and $n=N$
\begin{eqnarray}
   - \zeta_1   r_1(1)  && =    d_2 r_1(2)   -  (b_1 +d_1) r_1(1)  
   \nonumber \\
    - \zeta_1   r_1(N)  && =   b_{N-1} r_1(N-1)     -  d_N r_1(N)  
      \label{rightbd1n}
\end{eqnarray}
For the left eigenvector $l_1(.)$, the eigenvalues equations
read  in the bulk $n=2,..,N-1$
\begin{eqnarray}
   - \zeta_1   l_1(m)   =  \left[ l_1(m+1)-l_1(m)\right] b_m  +  \left[ l_1(m-1)-l_1(m)\right] d_m
\label{leftbd}
\end{eqnarray}
with the boundary equations for $n=1$ and $n=N$
\begin{eqnarray}
 - \zeta_1   l_1(1)    &&    =  \left[ l_1(2)-l_1(1)\right] b_1  +  \left[ 0 -l_1(1)\right] d_1 
   \nonumber \\
 - \zeta_1   l_1(N)    &&     =    \left[ l_1(N-1)-l_1(N)\right] d_N  
      \label{leftbd1n}
\end{eqnarray}

For arbitrary birth rates $b_n$ and death rates $d_n$, 
the solution of this spectral problem requires the introduction of orthogonal polynomials (see \cite{doorn} and references therein),
so that there is no simple explicit closed expression of the extinction rate $\zeta_1$ as a function of all the birth and death rates
of the model. However as described below, the extinction rate can be written explicitly in terms of the empirical observables of the metastable state, both at Level 2.5 and at Level 2.


\subsection{ Probability $P^{[2.5]living}_T[ \rho(.) ; q(.,.) ]  $ to remain living up to $T$ with given empirical density and empirical flows  }

For a very long dynamical trajectory $n(0 \leq t \leq T)$
of the population in the birth-death process of Eq. \ref{masterbd},
the empirical observables are the empirical time-averaged density of the population $n$
\begin{eqnarray}
 \rho_T( n ) \equiv \frac{1}{T} \int_0^T dt \ \delta_{n(t) , n }
\label{rhobd}
\end{eqnarray}
and the empirical time-averaged flows between population $n$ and population $n \pm 1$
\begin{eqnarray}
q_T(n\pm 1,n) \equiv  \frac{1}{T} \sum_{t \in [0,T] : n(t^+) \ne n(t^-)} \delta_{n(t^+),n \pm 1} \delta_{x(t^-),n} 
\label{flowbd}
\end{eqnarray}

The probability to remain confined in the set $\Omega^*=\{1,2..,N\}$ 
of living configurations with giving empirical observables $[ \rho(.) ; q(.,.) ]  $
follows the large deviation form of Eq. \ref{level2.5b}
\begin{eqnarray}
P^{[2.5]living}_T[ \rho(.) ; q(.,.) ] 
\oppropto_{T \to +\infty} C^{living}_{2.5}[ \rho(.) ; q(.,.) ]  e^{- T I^{living}_{2.5}[ \rho(.) ; q(.,.) ] }
\label{level2.5bd}
\end{eqnarray}
with the constraints obtained from Eq. \ref{c2.5b} 
\begin{eqnarray}
C^{living}_{2.5}[ \rho(.) ; q(.,.) ] 
&& = \delta \left(  \rho(0) \right) \delta \left(  \sum_{n=1 }^N \rho(n)- 1 \right) 
\delta \left( q(0,1)  \right)
\delta \left(  q(1,2) - q(2,1) \right)  
\delta \left(  q(N,N-1) - q(N-1,N) \right)  
\nonumber \\
&&
\left[ \prod_{n =2}^{N-1} \delta \left(  q(n,n+1)+q(n,n-1) - q(n+1,n)-q(n-1,n) \right)  \right]
\label{c2.5bd}
\end{eqnarray}
while the rate function of Eq. \ref{rate2.5b} reads
\begin{eqnarray}
 I^{living}_{2.5}[ \rho(.) ; q(.,.) ] 
&& =    d_1 \rho(1) 
+  \sum_{n=1}^{N-1} 
\left[ q(n+1,n)  \ln \left( \frac{ q(n+1,n)  }{  b_n  \rho(n) }  \right)  - q(n+1,n)  + b_n  \rho(n)  \right] 
\nonumber \\
&& +  \sum_{n=1}^{N-1} 
\left[ q(n,n+1)  \ln \left( \frac{ q(n,n+1)  }{  d_{n+1}  \rho(n+1) }  \right)  - q(n,n+1)  + d_{n+1}  \rho(n+1)  \right]  
\label{rate2.5bd}
\end{eqnarray}

The stationarity constraints of Eq. \ref{c2.5bd} for the empirical flows 
means that the total current $\left[ q(n,n+1) - q(n+1,n)\right] $ on the link between $n$ and $(n+1)$ 
cannot depend on $n$ and should vanish as a consequence of the boundary conditions at $n=1$ and $n=N$
\begin{eqnarray}
0  =   q(2,1) - q(1,2)= ... =  
 =  q(n+1,n) - q(n,n+1)
 = ... = q(N,N-1) - q(N-1,N)
 \label{jq}
\end{eqnarray}
It is thus convenient to replace the two flows on each link
by the new function $q_s$ of their middle-point $( n+1/2)$
\begin{eqnarray}
 q(n,n+1) = q(n+1,n) = q_s(n+1/2)  
 \label{qs}
\end{eqnarray}
in order to rewrite the level 2.5 of Eq. \ref{level2.5bd}
where the only remaining constraints involve the empirical density $\rho(.)$
\begin{eqnarray}
P^{[2.5]living}_T[ \rho(.) ; q_s(.) ] \oppropto_{T \to +\infty} 
 \delta \left(  \rho(0) \right)
 \delta \left(  \sum_{n=1}^N \rho(n)- 1 \right) 
 e^{- T I^{living}_{2.5}[ \rho(.) ; q_s(.) ] }
\label{level2.5bdqs}
\end{eqnarray}
while the corresponding rate function obtained from Eq. \ref{rate2.5bd} reads
\begin{eqnarray}
 I^{living}_{2.5}[ \rho(.) ; q_s(.) ] && = d_1 \rho(1) 
 \nonumber \\  &&
 +\sum_{n=1}^{N-1} 
\left[ q_s(n+1/2)  \ln \left( \frac{ q_s^2(n+1/2)  }{  b_n  \rho(n) d_{n+1}  \rho(n+1)}  \right) 
 - 2 q_s(n+1/2)  + b_n  \rho(n) + d_{n+1}  \rho(n+1) \right] 
\label{rate2.5bdqs}
\end{eqnarray}


\subsection{ Probability $P^{[2]living}_T[ \rho(.)  ]  $ to remain living up to $T$ with given empirical density   }

For the present model, the contraction explained around Eq. \ref{level2b} can be explicitly computed as follows.
For a given empirical density $\rho(.)$,
the optimization of the rate function of Eq. \ref{rate2.5bdqs}
with respect to the link flow $q_s(n+1/2)$
\begin{eqnarray}
&&  0 = \frac{ \partial  I^{living}_{2.5}[ \rho(.) ; q_s(.) ] }{\partial q_s(n+1/2) }
=   \ln \left( \frac{ q_s^2(n+1/2)  }{  b_n  \rho(n) d_{n+1}  \rho(n+1)}  \right) 
 \label{rate2.5qsderi}
\end{eqnarray}
yields the optimal value
\begin{eqnarray}
q^{opt}_s(n+1/2)  = \sqrt{ b_n  \rho(n) d_{n+1}  \rho(n+1)  }
 \label{rate2.5deri}
\end{eqnarray}
So the probability to remain living during $T$
  with the empirical density $\rho(.) $
follows the large deviation form of Eq. \ref{level2b}
\begin{eqnarray}
P_T^{[2]living} [ \rho(.) ]
  \opsimeq_{T \to +\infty}  \delta \left(  \rho(0) \right) \delta \left( \sum_{n=1}^N \rho(n)  - 1 \right) 
   e^{ - T I_2^{living}  [ \rho(.)] } 
\label{level2master}
\end{eqnarray}
where the rate function $I_{2}^{living}  [ \rho(.)] $ at Level 2 is obtained 
from the optimal value of the rate function at the Level 2.5 of Eq. \ref{rate2.5bdqs} 
\begin{eqnarray}
 I^{living}_{2}  [ \rho(.)] && = I_{2.5}[ \rho(.) ; q^{opt}_s(.) ] 
 \\
&&  = d_1 \rho(1)  + \sum_{n=1}^{N-1} 
\left[   b_n  \rho(n) + d_{n+1}  \rho(n+1) -2  \sqrt{ b_n  \rho(n) d_{n+1}  \rho(n+1)  }
 \right] 
\nonumber \\
&& = d_1 \rho(1)  + \sum_{n=1}^{N-1} 
\left[  \sqrt{ b_n  \rho(n) } -  \sqrt{  d_{n+1}  \rho(n+1)  } \right]^2
\label{rate2bd}
\end{eqnarray}


\subsection{ Probability $P^{living}_T  \propto_{T \to +\infty} e^{- T \zeta_1 } $ to remain living up to $T$    }

As explained in the subsection \ref{subsec_equivalence} for the general case, 
the exact optimization procedure
to obtain $\zeta_1$ governing the probability to remain living up to $T$ 
\begin{eqnarray}
P^{living}_T   \oppropto_{T \to +\infty} e^{- T \zeta_1 }
\label{level1zeta1bd}
\end{eqnarray}
via the contraction of the Level 2.5 is equivalent to the spectral problem of subsection \ref{subsec_spectralbd}
for the slowest relaxation mode.

As explained in the subsection \ref{subsec_zeta1perfromlargedev}, 
the estimation of $\zeta_1^{per}$ obtained via the perturbation theory in the absorbing rate $d_1^-$
can be recovered from the Level 2.5 via the procedure explained before Eq. \ref{level2.5bstar}.
Here we can obtain it directly from the Level 2 of Eq. \ref{level2master} as follows.
One just needs to require that all the bulk contributions of the rate function $I^{living}_{2}  [ \rho(.)] $ of Eq. \ref{rate2bd}
vanish
\begin{eqnarray}
  \sqrt{ b_n  \rho_*(n) } -  \sqrt{  d_{n+1}  \rho_*(n+1)  } =0 \ \ {\rm for } \ \ n=1,2,..,N-1
\label{rhostar}
\end{eqnarray}
The solution of this recurrence reads
\begin{eqnarray}
 \rho_*(n) =  \rho_*(1) \left( \prod_{j=1}^{n-1} \frac{ b_{j} }{d_{j+1}} \right)
\label{r0star}
\end{eqnarray}
where $  \rho_*(1) $ is determined by the normalization
\begin{eqnarray}
1=\sum_{n=1}^{N}   \rho_*(n) 
=  \rho_*(1) \left[ 1+ \sum_{n=2}^{N} \prod_{j=1}^{n-1} \frac{ b_{j} }{d_{j+1}} \right]
=  \rho_*(1) \left[ 1+ \frac{ b_1 }{d_2} +  \frac{ b_1 b_2 }{d_2 d_3} +... +  \frac{ b_1 b_2 ... b_{N-1} }{d_2 d_3 ... d_N } \right]
\label{z}
\end{eqnarray}
So the leading contribution 
for the slowest relaxation rate 
reduces to the flow entering the absorbing state $x=0$ from the neighboring site $y=1$
\begin{eqnarray}
\zeta_1^{per}  =   d_1 \rho_*(1)
=  \frac{d_1}{ \displaystyle 1+ \sum_{n=2}^{N} \prod_{j=1}^{n-1} \frac{ b_{j} }{d_{j+1}}  }
  \label{zetaperbd}
\end{eqnarray}
As mentioned after Eq. \ref{zeta1per}, this perturbative estimation will be small 
even if the absorbing rate $d_1 $ is not particularly small, as long as the empirical density
$ \rho_*(1)$ is small for population $n=1$, i.e. as long as the denominator of Eq. \ref{zetaperbd} involving all the other rates of the model is large.


\section{  Large deviations for the birth-death model in a switching environment }

\label{sec_twolevel}

Since population stochastic models in a randomly switching environment 
have attracted a lot of interest recently \cite{frey2017,frey2018,assaf2020},
this section is devoted to a simple birth-death model in an environment 
that can switch between two possible states $\sigma=\pm 1$.

\subsection{  Birth-death model in an environment switching between two possible states $\sigma=\pm 1$}

The two possible states $\sigma=\pm$ of the environment determine the dynamics for the population $n$ as follows.
In the positive environment $\sigma=+1$, only births are possible with positive birth rates $b_n>0 $ for population $n=1,..,N-1$.
In the negative environment $\sigma=-1$, only deaths are possible with positive death rates $d_n>0$ for $n=1,2,..,N$.
At population $n=1,..,N$, the environment state $\sigma$ can switch to the opposite value $(-\sigma)$
with the switching rate $\gamma_n^{\sigma} $ that may depend on the value $n$ of the population
if one wishes to describe some influence of the population on the environment.
The dynamics for the joint distribution $ P^{\sigma}_t(n)  $ to be at population $n =1,2,..,N$ and in the environment $\sigma=\pm$ 
is described by the master equations
\begin{eqnarray}
\partial_t  P^+_t(n)  && = b_{n-1} P^+_t(n-1) - b_n  P^+_t(n) 
- \gamma_n^+ P^+_t(n) +  \gamma_n^- P^-_t(n)
\nonumber \\
\partial_t   P^-_t(n) && = d_{n+1}  P^-_t(n+1) -  d_n P^-_t(n)  
+ \gamma_n^+  P^+_t(n) -  \gamma_n^-  P^-_t(n)
\label{bdpm}
\end{eqnarray}
while the dynamics of the dead configuration $n=0$ only involves the absorbing flow governed by the death rate $d_1^-$
\begin{eqnarray}
\partial_t  P_t(0)   =     d_1 P^-_t(1) 
\label{bdpm0}
\end{eqnarray}


\subsection{ Probability $P^{[2.5]living}_T[ \rho(.) ; q(.,.) ]  $ to remain living up to $T$ with given empirical density and empirical flows  }

For the dynamics of Eq. \ref{bdpm}, 
the probability at Level 2.5 of Eq. \ref{level2.5b}
\begin{eqnarray}
P^{[2.5]living}_T[ \rho^{\pm}(.) ; j^{\pm}(.) ; q^{\pm}(.) ] 
\oppropto_{T \to +\infty} C^{living}_{2.5}[ \rho^{\pm}(.) ; j^{\pm}(.) ; q^{\pm}(.)  ]  
e^{- T I^{living}_{2.5}[ \rho^{\pm}(.) ; j^{\pm}(.) ; q^{\pm}(.)  ] }
\label{level2.5bpm}
\end{eqnarray}
involve the following empirical time-averaged observables $[ \rho^{\pm}(.) ; j^{\pm}(.) ; q^{\pm}(.) ]$
for a long trajectory $[ n(0 \leq t \leq T); \sigma(0 \leq t \leq T)$ :

(i) The empirical densities of the population $n$ and of the environment $\sigma=\pm 1$
\begin{eqnarray}
 \rho^{ \sigma}( n)  \equiv \frac{1}{T} \int_0^T dt  \delta_{\sigma(t),\sigma}  \ \delta_{  n(t),  n} 
 \label{rho1def}
\end{eqnarray}
  satisfy the global normalization on the set of living configurations 
\begin{eqnarray}
  \sum_{ n=1 }^N  \left[ \rho^{ +}( n) + \rho^{ -}( n)\right]=1
\label{rho1norma}
\end{eqnarray}

(ii) The empirical current between populations $n$ and $(n+1)$ in the environment $\sigma=+$
\begin{eqnarray}
 j^+(n+1/2)  \equiv  \frac{1}{T} \sum_{t \in [0,T]: n(t^+) \ne n(t^-)} \delta_{\sigma(t),+}
 \delta_{n(t^+),n+1} \delta_{n(t^-),n} 
\label{jp}
\end{eqnarray}
and the empirical current between populations $n$ and $(n-1)$ in the environment $\sigma=-$
\begin{eqnarray}
 j^-(n-1/2)  \equiv  \frac{1}{T} \sum_{t \in [0,T]: n(t^+) \ne n(t^-)} \delta_{\sigma(t),-}
 \delta_{n(t^+),n-1} \delta_{n(t^-),n} 
\label{jpm}
\end{eqnarray}
are labelled by their middle-point $(n \pm 1/2)$ to simplify the notations.

(iii) At population $n$, the switching events between the two sates $\sigma=\pm 1$ of the environment
are described by the empirical switching flows  
\begin{eqnarray}
q^+( n)  \equiv  \frac{1}{T} \sum_{t  \in [0,T]: \sigma(t^+)=- \ne \sigma(t^-)=+ }  \  \delta_{n(t),  n} 
\nonumber \\
q^-( n)  \equiv  \frac{1}{T} \sum_{t  \in [0,T]: \sigma(t^+) =+\ne \sigma(t^-)=- }   \delta_{n(t),  n}
\label{jumpflows}
\end{eqnarray}
 The stationarity conditions mean that the total flow into the state 
$(n,\sigma)$ should be balanced by the total flow out of the state $(n,\sigma)$.
These conditions read for population $n=2,..,N-1$ and the two possible states $\sigma=\pm 1$ of the environment
\begin{eqnarray}
  0  &&  =    j^+(n-1/2)  - j^+(n+1/2)  - q^+(n)+q^-(n)
 \nonumber \\
0  &&  =    j^-(n+1/2)  - j^-(n-1/2)  + q^+(n)-q^-(n)
\label{statio}
\end{eqnarray}
with the boundary equations for $n=1$ and $n=N$ correspond to $ j^+(1/2)=0= j^+(N+1/2)$.
The sum of the two Eqs \ref{statio} yield that the total empirical current $\left[j^+(n+1/2) -  j^-(n+1/2) \right]$ vanishes for all $n$ 
\begin{eqnarray}
0  = j^+(3/2)-  j^-(3/2) = ...  = j^+(n+1/2) -  j^-(n+1/2) = ...  = j^+(N-1/2) -J^-(N-1/2)
 \label{jconserv}
\end{eqnarray}
As a consequence, it is possible to eliminate all the empirical negative 
currents in terms of the empirical positive currents
\begin{eqnarray}
j^-(n+1/2) = j^+(n+1/2)
 \label{jmjp}
\end{eqnarray}
while the remaining stationarity conditions involving the empirical switching flows $q^{\pm}(n)$ read
\begin{eqnarray}
    q^-(n) -   q^+(n)   =  j^+(n+1/2) -     j^+(n-1/2) 
     \label{jqst}
\end{eqnarray}

In summary, the large deviations of Eq. \ref{level2.5bpm} can be written without the empirical negative current $j^-(.) $ 
with the following constraints
\begin{eqnarray}
 C^{living}_{2.5}[ \rho^{\pm}(.) ; j^+(.) ; q^{\pm}(.)  ]  
  = \delta \left( \sum_{n=1}^N \left[ \rho^{ +}( n) + \rho^{ -}( n)\right] -1 \right)
 \prod_{n=1}^N \delta \left[ q^-(n) -   q^+(n)  -  j^+(n+1/2) +     j^+(n-1/2) \right]
\label{c2.5bpm}
\end{eqnarray}
and the following rate function at Level 2.5
\begin{eqnarray}
&& I_{2.5}^{living}[ \rho^{\pm}(.) ; j^+(.) ; q^{\pm}(.)  ]  
 = d_1 \rho^-(1)  +  \sum_{ n=1 }^{N-1}
\left[  j^+(n+1/2)   \ln \left( \frac{  j^+(n+1/2)  }{  b_n \rho^+(n)  }  \right) 
 -   j^+(n+1/2)  +    b_n \rho^+(n)     \right]  
\nonumber \\
&& +   \sum_{ n=1}^{N-1}
\left[  j^+(n+1/2)   \ln \left( \frac{ j^+(n+1/2)   }{ d_{n+1} \rho^-(n+1)  }  \right) 
 -  j^+(n+1/2)    + d_{n+1} \rho^-(n+1)  \right]  
\nonumber \\
&& 
 +  \sum_{ n=1}^N
\left[  q^+(n)    \ln \left( \frac{  q^+(n) }{ \gamma^+_n \rho^+(n)  }  \right) 
 -   q^+(n)   +  \gamma^+_n \rho^+(n)    \right]  
  +  \sum_{ n=1}^N
\left[  q^-(n)    \ln \left( \frac{  q^-(n) }{ \gamma^-_n \rho^-(n)  }  \right) 
 -   q^-(n)   +  \gamma^-_n \rho^-(n)    \right]  
\label{rate2.5bpm}
\end{eqnarray}


\subsection{ Explicit contraction over the switching activity $a(n) $ to obtain the intermediate Level 2.25  }

It is convenient to replace the two switching flows $q^{\pm}(n) $ by the switching activity $a(n) $ 
and the switching current $ i(n)$ representing their symmetric and antisymmetric parts
\begin{eqnarray}
a(n) \equiv  q^+(n) +   q^-(n)  
\nonumber \\
i(n) \equiv  q^+(n) -   q^-(n)  
 \label{ai}
\end{eqnarray}
i.e.
\begin{eqnarray}
 q^+(n) && = \frac{a(n)+i(n)}{2} 
\nonumber \\
q^-(n) && = \frac{a(n)-i(n)}{2} 
 \label{aiinv}
\end{eqnarray}

The constraints of Eq. \ref{c2.5bpm} involve the switching current $ i(n)$ but not the the switching activity $a(n) $
\begin{eqnarray}
 C^{living}[ \rho^{\pm}(.) ; j^+(.) ; i(.) ]  &&
  = \delta \left( \sum_{n=1}^N \left[ \rho^{ +}( n) + \rho^{ -}( n)\right] -1 \right)
 \prod_{n=1}^N \delta \left[ j^+(n-1/2)  -  j^+(n+1/2) -i(n)      \right]
\label{c2.5bpmi}
\end{eqnarray}
As a consequence, as in many other Markov jump processes \cite{maes_canonical,c_ring,c_interactions,c_detailed,c_jumpdiff},
the rate function obtained from Eq. \ref{rate2.5bpm}
\begin{eqnarray}
&& I_{2.5}^{living}[ \rho^{\pm}(.) ; j^+(.) ; i(.);a(.)  ]  
 = d_1 \rho^-(1) 
\nonumber \\
&&  +  \sum_{ n=1 }^{N-1}
\left[  j^+(n+1/2)   \ln \left( \frac{ \left[ j^+(n+1/2)\right]^2  }{  b_n \rho^+(n)d_{n+1} \rho^-(n+1) }  \right) 
 -   2j^+(n+1/2)  +    b_n \rho^+(n)      + d_{n+1} \rho^-(n+1)  \right]  
\nonumber \\
&& 
 +  \sum_{ n=1}^N
\left[  \frac{a(n)+i(n)}{2}     \ln \left( \frac{  a(n)+i(n) }{ 2 \gamma^+_n \rho^+(n)  }  \right) 
 +  \frac{a(n)-i(n)}{2}     \ln \left( \frac{  a(n)-i(n) }{ 2 \gamma^-_n \rho^-(n)  }  \right) 
 -   a(n)  +  \gamma^+_n \rho^+(n)   +  \gamma^-_n \rho^-(n)    \right]  
\label{rate2.5bpmia}
\end{eqnarray}
can be optimized over the switching activity $a(n) $
\begin{eqnarray}
0 = \frac{ \partial   I_{2.5}^{living}[ \rho^{\pm}(.) ; j^+(.) ; i(.);a(.)  ]    }{ \partial a(n) }
 =   \frac{1}{2}     \ln \left( \frac{  a^2(n)- i^2(n) }{ 4 \gamma^+_n \rho^+(n)   \gamma^-_n \rho^-(n)  } \right)
  \label{rate2.5bpmiaderi}
\end{eqnarray}
in order to obtain the optimal value $ a^{opt}(n) $ as a function of the other empirical observables
\begin{eqnarray}
 a^{opt}(n) = \sqrt{  i^2(n) + 4 \gamma^+_n \rho^+(n)   \gamma^-_n \rho^-(n) }
  \label{actiopt}
\end{eqnarray}
Plugging this optimal value into the rate function at Level 2.5 of Eq. \ref{rate2.5bpmia}
yields the rate function at Level 2.25
\begin{eqnarray}
&&  I_{2.25}^{living}[ \rho^{\pm}(.) ; j^+(.) ; i(.)  ] =   I_{2.5}^{living}[ \rho^{\pm}(.) ; J^{-}(.) ; i(.);a^{opt}(.)  ]  
 = d_1 \rho^-(1) 
 \label{rate2.25}
 \\
&&  +  \sum_{ n=1 }^{N-1}
\left[  j^+(n+1/2)   \ln \left( \frac{ \left[ j^+(n+1/2)\right]^2  }{  b_n \rho^+(n)d_{n+1} \rho^-(n+1) }  \right) 
 -   2j^+(n+1/2)  +    b_n \rho^+(n)      + d_{n+1} \rho^-(n+1)  \right]  
\nonumber \\
&& 
 +  \sum_{ n=1}^N
\left[ i(n)     \ln \left( \frac{  \sqrt{  i^2(n) + 4 \gamma^+_n \rho^+(n)   \gamma^-_n \rho^-(n) }+i(n) }{ 2 \gamma^+_n \rho^+(n)  } \right) 
   +  \gamma^+_n \rho^+(n)   +  \gamma^-_n \rho^-(n)  -  \sqrt{  i^2(n) + 4 \gamma^+_n \rho^+(n)   \gamma^-_n \rho^-(n) }  \right]  
\nonumber
\end{eqnarray}
that will govern  
the probability to remain living up to $T$ with the empirical observables $[ \rho^{\pm}(.) ; j^+(.) ; i(.) ] $
\begin{eqnarray}
P^{[2.25]living}_T[ \rho^{\pm}(.) ; j^+(.) ; i(.) ] 
&& \oppropto_{T \to +\infty} e^{- T  I_{2.25}^{living}[ \rho^{\pm}(.) ; j^+(.) ; i(.)  ] }
\nonumber \\
&&  \delta \left( \sum_{n=1}^N \left[ \rho^{ +}( n) + \rho^{ -}( n)\right] -1 \right)
 \prod_{n=1}^N \delta \left[ j^+(n-1/2)  -  j^+(n+1/2) -i(n)      \right]
\label{level2.25bpm}
\end{eqnarray}
One can now use the last constraint to eliminate the switching currents $i(n) =\left[ j^+(n-1/2)  -  j^+(n+1/2)     \right]  $
and one obtains 
that the probability to remain living up to $T$ with the empirical observables $[ \rho^{\pm}(.) ; j^+(.)  ] $
only involves the normalization contraint for the empirical density
\begin{eqnarray}
P^{[2.25]living}_T[ \rho^{\pm}(.) ; j^+(.)  ] 
&& \oppropto_{T \to +\infty}  \delta \left( \sum_{n=1}^N \left[ \rho^{ +}( n) + \rho^{ -}( n)\right] -1 \right) 
e^{- T  I_{2.25}^{living}[ \rho^{\pm}(.) ; j^+(.)  ] }
\label{level2.15bpm}
\end{eqnarray}
while the corresponding rate function reads
\begin{footnotesize}
\begin{eqnarray}
&&  I_{2.25}^{living}[ \rho^{\pm}(.) ; j^+(.)  ] 
 = d_1 \rho^-(1) 
 \label{rate2.25jp}
 \\
&&  +  \sum_{ n=1 }^{N-1}
\left[  j^+(n+1/2)   \ln \left( \frac{ \left[ j^+(n+1/2)\right]^2  }{  b_n \rho^+(n)d_{n+1} \rho^-(n+1) }  \right) 
 -   2j^+(n+1/2)  +    b_n \rho^+(n)      + d_{n+1} \rho^-(n+1)  \right]  
\nonumber \\
&& 
 +  \sum_{ n=1}^N
\left[ \left[ j^+(n-1/2)  -  j^+(n+1/2)     \right]    \ln \left( \frac{  \sqrt{  \left[ j^+(n-1/2)  -  j^+(n+1/2)     \right]^2 + 4 \gamma^+_n \rho^+(n)   \gamma^-_n \rho^-(n) }+\left[ j^+(n-1/2)  -  j^+(n+1/2)     \right] }{ 2 \gamma^+_n \rho^+(n)  } \right) \right]
\nonumber \\
&& 
 +  \sum_{ n=1}^N
\left[     \gamma^+_n \rho^+(n)   +  \gamma^-_n \rho^-(n)  -  \sqrt{ \left[ j^+(n-1/2)  -  j^+(n+1/2)     \right]^2 + 4 \gamma^+_n \rho^+(n)   \gamma^-_n \rho^-(n) }  \right]  
\nonumber
\end{eqnarray}
\end{footnotesize}

To obtain the probability to remain living up to $T$ with the empirical densities $[ \rho^{\pm}(.) ] $ only,
one needs to integrate Eq. \ref{level2.15bpm} over the empirical current $ j^+(.)$
\begin{eqnarray}
P^{living}_T[ \rho^{\pm}(.) ] 
&& \oppropto_{T \to +\infty}  \delta \left( \sum_{n=1}^N \left[ \rho^{ +}( n) + \rho^{ -}( n)\right] -1 \right) 
\int {\cal D} j^+(.)
e^{- T  I_{2.25}^{living}[ \rho^{\pm}(.) ; j^+(.)  ] }
\nonumber \\
&& \oppropto_{T \to +\infty}  \delta \left( \sum_{n=1}^N \left[ \rho^{ +}( n) + \rho^{ -}( n)\right] -1 \right) 
e^{- T  I_{2}^{living}[ \rho^{\pm}(.)   ] }
\label{level2bpm}
\end{eqnarray}
The optimization of the rate function $I_{2.25}^{living}[ \rho^{\pm}(.) ; j^+(.)  ] $ 
at Level 2.25 of Eq. \ref{rate2.25jp} over the current $ j^+(.)$
needed to obtain the rate function $I_{2}^{living}[ \rho^{\pm}(.)   ]  $ at Level 2 
is somewhat heavy and will not be discussed further.

As a final remark, let us mention 
that the continuous-space analog of the birth-death model in a switching environment of Eq. \ref{bdpm}
corresponds to the run-and-tumble process with space-dependent velocities $v_{\pm}(x)$ 
and space-dependent switching rates $\gamma_{\pm}(x)$ : 
their large deviations properties at various levels studied recently for the case without absorption \cite{c_runandtumble} are somewhat simpler in the continuous model
because the empirical currents $j^{\pm}(x)$ are then completely 
determined by the empirical densities $\rho^{\pm}( x) $ 
as a consequence of the deterministic motion at velocities $v_{\pm}(x)$ when
the environment is state $(\pm)$.


\subsection{ Probability $P^{living}_T  $ to remain living up to $T$    }

As explained in the subsection \ref{subsec_equivalence}, the exact optimization procedure
to obtain $\zeta_1$ via the contraction of the Level 2.5 is equivalent to the spectral problem 
for the slowest relaxation mode described in section \ref{sec_spectral}.
So here we will only discuss how
the estimation of $\zeta_1^{per}$ of the perturbation theory in the absorbing rate $d_1^-$
can be obtained directly from the Level 2.5 via the procedure explained before Eq. \ref{level2.5bstar}.
To obtain the steady state on the set $\Omega^*$ of living configurations,
 one requires the vanishing of all the bulk contributions of 
the rate function at Level 2.5 of Eq. \ref{rate2.5bpm}
\begin{eqnarray}
 j^+_*(n+1/2) && = b_n \rho_*^+(n)   \ \ \ \ \  \ \ \ \ \ \ \ \ \ \ \ \ {\rm for } \ \ \ n=1,..,N-1
  \nonumber \\ 
 j^+_*(n+1/2)   &&= d_{n+1} \rho_*^-(n+1)    \ \ \ \ \  \ \ \ {\rm for } \ \ \ n=1,..,N-1
  \nonumber \\ 
  q_*^+(n) && = \gamma^+_n \rho_*^+(n)   \ \ \ \ \  \ \ \ \ \ \ \ \ \ \ \ {\rm for } \ \ \ n=1,..,N
   \nonumber \\  
    q^-_*(n)  && =  \gamma^-_n \rho_*^-(n)   \ \ \ \ \  \ \ \ \ \ \ \ \ \ \ \ {\rm for } \ \ \ n=1,..,N
\label{rate2.5bpmtyp}
\end{eqnarray}
and the satisfaction of all the constraints of Eq. \ref{c2.5bpm}
\begin{eqnarray}
1 && = \sum_{n=1}^N \left[ \rho_*^{ +}( n) + \rho_*^{ -}( n)\right] 
 \nonumber \\
 0 && =  q^-_*(n) -   q^+_*(n)  -  j^+_*(n+1/2) +     j^+_*(n-1/2) \ \ {\rm for } \ \ \ n=2,..,N-1
  \nonumber \\
 0 && =  q^-_*(1) -   q^+_*(1)  -  j^+_*(3/2) 
  \nonumber \\
 0 && =  q^-_*(N) -   q^+_*(N)   +     j^+_*(N-1/2)
\label{c2.5bpmtyp}
\end{eqnarray}
One can use the two first equations of Eq. \ref{rate2.5bpmtyp}
to rewrite the densities $\rho^{\pm}_*(.)$ in terms of the positive steady current $j^+_*(.) $
\begin{eqnarray}
\rho_*^{+}(n) && = \frac{ j^+_*(n+1/2) }{ b_n  } \ \ {\rm for } \ \ n=1,2,..,N-1
 \nonumber \\
\rho_*^-(n) && =  \frac{ j^+_*(n-1/2) }{d_n } \ \ {\rm for } \ \ n=2,2,..,N
\label{pfromj}
\end{eqnarray}
Plugging the two last equations of Eq. \ref{rate2.5bpmtyp} for $n=1$ and $n=N$
into the two last equations of Eq. \ref{c2.5bpmtyp}
allow to compute the two missing densities  
\begin{eqnarray}
  \rho^-_*(1) && =  j^+_*(3/2) \frac{ 1+  \frac{ \gamma_1^+ }{b_1^+} }{ \gamma_1^-}
\nonumber \\
 \rho^+_*(N) && = j^+_*(N-1/2) \frac{ 1  + \frac{ \gamma_N^- }{d_N} }{ \gamma_N^+}
\label{bdpmstnp}
\end{eqnarray}
Plugging the two last equations of Eq. \ref{rate2.5bpmtyp} for $n=2,..,N-1$
into the second equation of Eq. \ref{c2.5bpmtyp}
 allows to compute the empirical current $ j^+_*(n+1/2)  $
on all the links 
\begin{eqnarray}
 j^+_*(n+1/2)    =  j^+_*(n-1/2)  \left[ \frac{ 1+  \frac{ \gamma_n^- }{d_n } }{1+ \frac{ \gamma_n^+ }{ b_n  } } \right]
 = ...
 =  j^+_*(3/2) \prod_{j=2}^n \left[ \frac{ 1+  \frac{ \gamma_j^- }{d_j } }{1+ \frac{ \gamma_j^+ }{ b_j  } } \right]
 \ \ {\rm for } \ \ n=2,..,N-1
\label{bdpmstj}
\end{eqnarray}
in terms of $ j^+_*(3/2)$ that should now be computed from the normalization of the total density 
\begin{eqnarray}
&& 1  = \sum_{n=1}^{N-1} \rho^+_*(n) + \rho_*^+(N) +   \rho^-_*(1) +  \sum_{n=2}^{N} \rho_*^-(n)
\label{normapm} \\
&& = j^+_*(3/2) \frac{ 1+  \frac{ \gamma_1^+ }{b_1^+} }{ \gamma_1^-}
+  \sum_{n=1}^{N-1} \left[ \frac{ 1 }{ b_n  } + \frac{ 1 }{d_{n+1}^- }\right] j^+_*(n+1/2) 
+  j^+_*(N-1/2) \frac{ 1  + \frac{ \gamma_N^- }{d_N} }{ \gamma_N^+}
\nonumber \\
&& = j^+_*(3/2) \left[ \frac{ 1+  \frac{ \gamma_1^+ }{b_1^+} }{ \gamma_1^-}
+  \frac{ 1 }{ b_1^+  } + \frac{ 1 }{d_2^- }\right]
+  \sum_{n=2}^{N-2} \left[ \frac{ 1 }{ b_n  } + \frac{ 1 }{d_{n+1}^- }\right] j^+_*(n+1/2) 
+  j^+_*(N-1/2) \left[\frac{ 1 }{ b_{N-1}  } + \frac{ 1 }{d_N} + \frac{ 1  + \frac{ \gamma_N^- }{d_N} }{ \gamma_N^+}
\right]
\nonumber
\end{eqnarray}
The solution of Eq. \ref{bdpmstj}
leads to the final result for $ j^+_*(3/2) $ 
\begin{eqnarray}
 \frac{1 }{  j^+_*(3/2) }
&& = \left( \frac{ 1+  \frac{ \gamma_1^+ }{b_1^+} }{ \gamma_1^-}+  \frac{ 1 }{ b_1^+  } + \frac{ 1 }{d_2^- }\right)
+  \sum_{n=2}^{N-2} \left( \frac{ 1 }{ b_n  } + \frac{ 1 }{d_{n+1}^- }\right)
\prod_{j=2}^n \left[ \frac{ 1+  \frac{ \gamma_j^- }{d_j } }{1+ \frac{ \gamma_j^+ }{ b_j  } } \right]
 \nonumber \\ &&
+ \left(\frac{ 1 }{ b_{N-1}  } + \frac{ 1 }{d_N } + \frac{ 1  + \frac{ \gamma_N^- }{d_N} }{ \gamma_N^+}
\right) \prod_{j=2}^{N-1} \left[ \frac{ 1+  \frac{ \gamma_j^- }{d_j } }{1+ \frac{ \gamma_j^+ }{ b_j  } } \right]
\label{j32}
\end{eqnarray}
So the perturbative evaluation 
for the slowest relaxation rate 
involves the probability $\rho^-_*(1)  $ of Eq. \ref{bdpmstnp}
\begin{eqnarray}
\zeta_1^{per}  =   d_1^- \rho^-_*(1) =   \frac{  d_1^- }{ \gamma_1^-} 
\left(1+  \frac{ \gamma_1^+ }{b_1^+} \right)  j^+_*(3/2)
  \label{zetaperbdpm}
\end{eqnarray}
where $ j^+_*(3/2)$ has been written in Eq. \ref{j32}
in terms of all the rates of the model.
This perturbative estimation $\zeta_1^{per} $
will be small even if the absorbing rate $d_1^-$ is not particularly small, if the empirical current $ j^+_*(3/2) $ is small
i.e. if its inverse $\frac{1}{j^+_*(3/2)}$ written in Eq. \ref{j32} is large.


\section{ Conclusion  }

\label{sec_conclusion}

In this paper, we have applied the large deviations at Level 2.5 to Markov processes with absorbing states
in order to obtain the explicit extinction rate of metastable quasi-stationary states
 in terms of their empirical time-averaged density and of 
 their empirical time-averaged flows over a large time-window $T$.
The case of Markov jump processes has been analyzed in detail in the main text,
 while the adaptation to diffusion processes in dimension $d$ iss described in Appendix \ref{app_fp}.
In both cases, we have explained how the full optimization of the extinction rate at Level 2.5 over all empirical observables allows to recover the standard spectral problem for the slowest relaxation mode
and we have discussed the link with the Doob generator of the process conditioned to survive up to time $T$.
Finally, this general formalism has been illustrated with the application to the population birth-death model 
in a stable or in a switching environment.

Our main conclusion is thus that the large deviations at Level 2.5 
that have been introduced to characterize the possible
dynamical fluctuations of non-equilibrium steady states of Markov processes
directly provide the appropriate framework  
to analyze the fluctuations properties of metastable quasi-stationary states.


\appendix

\section{ Adaptation to diffusion processes in dimension $d$ with absorbing states  }

\label{app_fp} 

In this Appendix, we consider diffusion processes  described by the Fokker-Planck dynamics
with drift $\vec v(\vec x)$ and diffusion coefficient $D(\vec x)$ in dimension $d$
\begin{eqnarray}
 \partial_t P_t(\vec x)     =  -   \vec \nabla .  \left[ P_t(\vec x )   \vec v(\vec x ) 
-D (\vec x) \vec \nabla   P_t(\vec x)  \right] \equiv {\cal F} P_t(\vec x )
\label{fokkerplanck}
\end{eqnarray}
on the set of  living configurations $\vec x \notin A$, while $A$ respresent the set of absorbing states.
The goal is to describe how the analysis of the main text concerning Markov jump processes
should be adapted when the Markov matrix $w$ is replaced by the Fokker-Planck 
differential operator $ {\cal F} $ of Eq. \ref{fokkerplanck}.


\subsection{ Reminder on the spectral analysis of the slowest relaxation mode before extinction }

\label{app_spectral}

The eigenvalues Eqs \ref{eigen1living} on the set of living configurations $\vec x \notin A$
have to be adapted as follows :
the equation for the positive right eigenvector $r_1(\vec x)$ for $\vec x \notin A$
involves the Fokker-Planck generator $ {\cal F} $
of Eq. \ref{fokkerplanck} 
\begin{eqnarray}
-\zeta_1 r_1(\vec x)   ={\cal F} r_1(\vec x) =   -   \vec \nabla .  \left[ r_1(\vec x )   \vec v(\vec x ) \right]
+ \vec \nabla .  \left[  D (\vec x) \vec \nabla   r_1(\vec x)  \right] 
\label{fokkerplanckright}
\end{eqnarray}
while the equation for the positive left eigenvector $l_1(\vec x)$ for $\vec x \notin A$
involves the adjoint operator $ {\cal F}^{\dagger} $
 \begin{eqnarray}
-\zeta_1 l_1(\vec x)   ={\cal F}^{\dagger} l_1(\vec x) =     \vec v(\vec x ) .  \vec \nabla  l_1(\vec x)
+ \vec \nabla .  \left[  D (\vec x) \vec \nabla   l_1(\vec x)  \right] 
\label{fokkerplanckleft}
\end{eqnarray} 
The Doob generator that has the probability distribution of Eq. \ref{interiorconditioned}
\begin{eqnarray}
\pi^{interior}(\vec x) =  l_1(\vec x)    r_1(\vec x)  
\label{interiorconditionedfp}
\end{eqnarray}
as steady state is the generalization of Eq. \ref{doob} given by the differential operator
\begin{eqnarray}
{ \hat {\cal F} } =  l_1(.)  {\cal F} \frac{1}{l_1(.) } + \zeta_1
\label{doobfp}
\end{eqnarray}
One obtains that the process conditioned to survive satisfy the modified Fokker-Planck dynamics
 \begin{eqnarray}
 \partial_t {\hat P}_t(\vec x)    =  { \hat {\cal F} } {\hat P}_t(\vec x )
= -   \vec \nabla .  \left[ {\hat P}_t(\vec x )   \vec u(\vec x ) 
-D (\vec x) \vec \nabla   {\hat P}_t(\vec x )  \right]
\label{fokkerplanckdoob}
\end{eqnarray}
where the only change with respect to the initial dynamics of Eq. \ref{fokkerplanck}
is that the initial drift $ \vec v(\vec x ) $ has been replaced by the effective drift
\begin{eqnarray}
\vec u (\vec x )   \equiv \vec v(\vec x )  +2D(\vec x)   \frac{ \vec \nabla  l_1(\vec x) }{ l_1(\vec x)}
\label{udoob}
\end{eqnarray}
Since the left eigenvector $ l_1(\vec x)$ vanishes on $A$ while it is positive for $x \notin A$,
this effective drift $\vec u (\vec x ) $ becomes very large near the surface of $A$ 
and prevents the conditioned process to reach $A$.

For more details about this spectral analysis of the slowest mode
and the corresponding conditioned Doob process for diffusion processes in dimensions $d=1,2,3$ 
with explicit examples, we refer to the recent works \cite{killing,Mazzolo_Taboo} and references therein.


\subsection{ Large deviations at various levels for metastable quasi-stationary states }


\subsubsection{ Reminder on the large deviations at Level 2.5 for the time-averaged density and the time-averaged current  }

For a very long dynamical trajectory $\vec x(0 \leq t \leq T)$ of the Fokker-Planck dynamics
of Eq. \ref{fokkerplanck}, the joint distribution of
the empirical density
\begin{eqnarray}
 \rho(\vec x) && \equiv \frac{1}{T} \int_0^T dt \  \delta^{(d)} ( \vec x(t)- \vec x)  
\label{rhodiff}
\end{eqnarray}
and of the empirical current $\vec j(\vec x)$
\begin{eqnarray} 
\vec j(\vec x) \equiv   \frac{1}{T} \int_0^T dt \ \frac{d \vec x(t)}{dt}   \delta^{(d)}( \vec x(t)- \vec x)  
\label{diffjlocal}
\end{eqnarray}
satisfy the large deviation form \cite{wynants_thesis,maes_diffusion,chetrite_formal,engel,chetrite_HDR,c_reset,c_lyapunov,c_inference}
\begin{eqnarray}
 P^{[2.5]}_T[ \rho(.), \vec j(.)]   \opsimeq_{T \to +\infty}  C_{2.5} [ \rho(.), \vec j(.)] 
 e^{- \displaystyle T I_{2.5} [ \rho(.), \vec j(.)] }
\label{level2.5diff}
\end{eqnarray}
where the constitutive constraints 
\begin{eqnarray}
C_{2.5} [ \rho(.), \vec j(.)]  =
 \delta \left(\int d^d \vec x \rho(\vec x) -1  \right)
\left[ \prod_{\vec x }  \delta \left(  \vec \nabla . \vec j(\vec x) \right) \right]
\label{c2.5diff}
\end{eqnarray}
contains the normalization of the empirical density $\rho(.)$ and 
the stationarity constraint given by
the divergence-free property for the empirical current $\vec j (.) $,
while the rate function is explicit for any Fokker-Planck dynamics in terms of the drift $\vec v(\vec x) $ and 
of the diffusion coefficient $D(\vec x) $
\begin{eqnarray}
I_{2.5} [ \rho(.), \vec j(.)]  =
\int \frac{d^d \vec x}{ 4 D(\vec x) \rho(\vec x) } \left[ \vec j(\vec x) - \rho(\vec x) \vec v(\vec x)+D(\vec x) \vec \nabla \rho(\vec x) \right]^2 
\label{rate2.5diff}
\end{eqnarray}


\subsubsection{ Probability $P^{[2.5]living}_T[ \rho(.) ; \vec j(.) ]  $ to remain living up to $T$ with given empirical density and empirical current  }

To obtain from Eq. \ref{level2.5diff} the probability to remain living up to time $T$
with given empirical observables, one needs to impose that the empirical density vanishes on 
the set $A$ of absorbing sites
\begin{eqnarray}
\rho(\vec x) && =0 \ \ {\rm for  } \ \ \vec x \in A
\label{0surA}
\end{eqnarray}
and that the empirical current entering the set $A$ of absorbing states vanishes,
i.e. using the normal vector $\vec n_{A}$ to the surface $\partial A$ of $A$
\begin{eqnarray}
\vec j (\vec x) . \vec n_A&& =0 \ \ {\rm for  } \ \ \vec x \in \partial A
\label{jenter0surA}
\end{eqnarray}
So the probability $P^{[2.5]living}_T[ \rho(.) ; \vec j(.) ]  $ to remain living up to $T$ with given empirical density and empirical current 
\begin{eqnarray}
 P^{[2.5]living}_T[ \rho(.), \vec j(.)]   \opsimeq_{T \to +\infty}  C^{living}_{2.5} [ \rho(.), \vec j(.)] 
 e^{- \displaystyle T I^{living}_{2.5} [ \rho(.), \vec j(.)] }
\label{level2.5diffb}
\end{eqnarray}
involve the constraints 
\begin{eqnarray}
C^{living}_{2.5} [ \rho(.), \vec j(.)]  = 
\left[ \prod_{ \vec x \in A} \delta \left( \rho(\vec x)   \right) \right]
\left[ \prod_{\vec x \in \partial A}  \delta \left(  \vec j (\vec x) . \vec n_A \right) \right]
 \delta \left(\int_{\vec x \notin A} d^d \vec x \rho(\vec x) -1  \right) 
\left[ \prod_{\vec x \notin A}  \delta \left(  \vec \nabla . \vec j(\vec x) \right) \right]
\label{c2.5diffb}
\end{eqnarray}
and the rate function 
\begin{eqnarray}
I_{2.5}^{living} [ \rho(.), \vec j(.)]  =
\int_{\vec x \notin A} \frac{d^d \vec x}{ 4 D(\vec x) \rho(\vec x) } \left[ \vec j(\vec x) - \rho(\vec x) \vec v(\vec x)+D(\vec x) \vec \nabla \rho(\vec x) \right]^2 
\label{rate2.5diffb}
\end{eqnarray}


\subsubsection{ Equivalence between the exact optimization of the Level 2.5 
and the spectral problem of subsection \ref{app_spectral}}

As in Eq. \ref{level1bfrom2.5}, 
the probability $P^{living}_T \propto e^{- T \zeta_1} $  to remain living up to $T$
that involves the slowest relaxation mode $\zeta_1$ discussed subsection \ref{app_spectral}
can be also computed via the integration of the Level 2.5 of Eq. \ref{level2.5diffb}
over the empirical density $\rho(.)$ and the empirical current $\vec j(.)$
\begin{eqnarray}
P^{living}_T && = \int {\cal D} \rho(.)  \int {\cal D} \vec j(.) P^{[2.5]living}_T[ \rho(.) ; \vec j(.) ] 
\label{level1bfrom2.5diff}
\end{eqnarray}
In order to solve this optimization problem,
let us introduce the following Lagrangian for $\vec x \notin A$
with the Lagrange multipliers $(\omega,\nu(.))$ 
to take into account the constraints of normalization and stationarity
\begin{eqnarray}
&& {\cal L}[ \rho(.),  \vec j(.) ] =
 \int_{\vec x \notin A} \frac{d^d \vec x}{ 4 D(\vec x) \rho(\vec x) } \left[ \vec j(\vec x) - \rho(\vec x) \vec v(\vec x)+D(\vec x) \vec \nabla \rho(\vec x) \right]^2 
  -  \omega \left( \int_{\vec x \notin A} d^d \vec x \rho(\vec x) -1  \right)
 + \int_{\vec x \notin A} d^d \vec x \nu(\vec x) \vec \nabla . \vec j(\vec x) 
 \nonumber \\
&& = \int_{\vec x \notin A} \frac{d^d \vec x}{ 4 D(\vec x) \rho(\vec x) } \left[ \vec j(\vec x) - \rho(\vec x) \vec v(\vec x)+D(\vec x) \vec \nabla \rho(\vec x) \right]^2 
  -  \omega \left( \int_{\vec x \notin A} d^d \vec x \rho(\vec x) -1  \right)
 - \int_{\vec x \notin A} d^d \vec x  \vec j(\vec x) .  \vec \nabla \nu(\vec x)
 \label{lagrangiandiff}
\end{eqnarray}
While the optimization can be written with these variables (see \cite{c_lyapunov} for instance),
it is technically simpler (as in the analog Eq. \ref{wempirical} for Markov jump processes)
to replace the empirical current $\vec j(\vec x)$
by the effective empirical drift $\vec u(\vec x)  $ computed from the empirical current and the empirical density
\begin{eqnarray}
\vec u(\vec x)  \equiv  \frac{\vec j (\vec x) + D(\vec x)   \vec \nabla \rho(\vec x)}{\rho(\vec x)} 
\label{uefflag}
\end{eqnarray}
After some integration by parts for the last term to eliminate the gradient of the empirical density, the Lagrangian of Eq. \ref{lagrangiandiff} translates into
\begin{eqnarray}
&& {\tilde {\cal L}}[ \rho(.),  \vec u(.) ] = \omega
+ \int_{\vec x \notin A} d^d \vec x \rho(\vec x)
\left[ \frac{\left[ \vec u(\vec x) - \vec v(\vec x) \right]^2}{ 4 D(\vec x)  }  
  -  \omega - \vec u(\vec x)   .  \vec \nabla \nu(\vec x)
    -  \vec \nabla \left(  D(\vec x)  \vec \nabla \nu(\vec x) \right)
   \right]
 \label{lagrangiandiffu}
\end{eqnarray}
The optimization with respect to the optimal drift $u(\vec x)$ leads to the optimal value
\begin{eqnarray}
\vec u^{opt} (\vec x) = \vec v(\vec x) + 2 D(\vec x)\vec \nabla \nu(\vec x)
 \label{uopt}
\end{eqnarray}
The optimization of the Lagrangian with respect to the empirical density $   \rho(\vec x)  $ yields
for this optimal drift the following closed equation for the Lagrange multiplier $\nu(.)$
\begin{eqnarray}
-  \omega && =  - \frac{\left[ \vec u^{opt}(\vec x) - \vec v(\vec x) \right]^2}{ 4 D(\vec x)  }  
   + \vec u^{opt}(\vec x)   .  \vec \nabla \nu(\vec x)
    +  \vec \nabla \left(  D(\vec x)  \vec \nabla \nu(\vec x) \right)
    \nonumber \\
&& =    D(\vec x)\left[  \vec \nabla \nu(\vec x) \right]^2
   +  \vec v(\vec x)   .  \vec \nabla \nu(\vec x)
    +  \vec \nabla \left(  D(\vec x)  \vec \nabla \nu(\vec x) \right)    
 \label{rhouopt}
\end{eqnarray}
The optimal value of the Lagrangian of Eq. \ref{lagrangiandiffu} that determines the slowest relaxation rate $ \zeta_1$ 
reduces to the Lagrange multiplier $\omega$ 
\begin{eqnarray}
\zeta_1 = {\hat {\cal L}}^{opt} = \omega 
\label{lagrangianoptdiff}
\end{eqnarray}

The comparison of the optimal drift of Eq. \ref{uopt} and the Doob drift of Eq. \ref{udoob}
leads to the change of variables
\begin{eqnarray}
\vec \nabla \nu(\vec x) =  \frac{ \vec \nabla  l_1(\vec x) }{ l_1(\vec x)}
\label{nul1}
\end{eqnarray}
that transforms indeed Eq. \ref{rhouopt} into the spectral eigenvalue equation of Eq. \ref{fokkerplanckleft}
for the left eigenvector $l_1(.)$.


\subsubsection{ Large deviations for general time-additive observables of the trajectory $\vec x(0 \leq t \leq T)$ before extinction  }

As explained in subsection \ref{subsec_additive} for Markov jump processes,
 the large deviations at Level 2.5 
allows to derive the large deviations of any time-additive observable of the Markov trajectory
using their decomposition in terms of empirical observables.
Here the empirical density $\rho(.)$ allows to 
reconstruct any time-additive observable ${\cal A}_T$ that involves some function $\alpha(\vec x)$  
\begin{eqnarray}
 {\cal A}_T   \equiv \frac{1}{T} \int_0^T dt \ \alpha(\vec x(t) )   
 = \int d^d \vec x \ \alpha(\vec x) \ \rho(\vec x)
\label{additiveAdiff}
\end{eqnarray}
while the empirical current $\vec j(.)$
allows to reconstruct any time-additive observable ${\cal B}_T$ that involves some function $\vec \beta(\vec x)$ 
\begin{eqnarray}
 {\cal B}_T  \equiv  \frac{1}{T}\int_0^T dt \ \frac{d \vec x(t)}{dt}
. \vec\beta( \vec x(t))  = \int d^d \vec x \   \vec j ( \vec x ) . \vec \beta (\vec x )
\label{additiveBdiff}
\end{eqnarray}
As a consequence, the large deviations at Level 2.5 for $P^{[2.5]living}_T[ \rho(.) ; \vec j(.) ]  $  of Eq. \ref{level2.5diffb}
allows to analyze the large deviations of any time-additive observable of the form $({\cal A}_T+{\cal B}_T) $
for trajectories living up to time $T$.


\subsection{ Application to the continuous-space analog of the birth-death model of section \ref{sec_bd} }

When the discrete population $n$ of the birth-death Markov jump process of Eq. \ref{masterbd}
is replaced by the continuous variable $x$, one is led to consider the Fokker-Planck dynamics
with drift $v(x)$ and diffusion coefficient $D(x)$
for $x>0$ 
\begin{eqnarray}
 \partial_t P_t( x)     =  -   \partial_x  \left[ P_t( x )   v( x ) 
-D (x)  \partial_x P_t( x)  \right] 
\label{fokkerplanck1d}
\end{eqnarray}
with absorption at the origin $x=0$, and reflecting boundary condition at the other boundary $x=N$.
Again, for the spectral analysis of the slowest mode and the corresponding conditioned Doob process  
with explicit examples, we refer to the recent works \cite{killing,Mazzolo_Taboo} and references therein.

For this one-dimensional interval $ x \in ]0,N]$, 
the current divergence-free constraint $\partial_x j(x)=0$ 
and the boundary conditions of zero current at the two boundary $x=0$ and $x=N$
yield that the empirical current identically vanishes $j(x)=0$.
So the Level 2.5 of Eq. \ref{level2.5diffb}
directly reduces to the Level 2 involving only the empirical density $\rho(.)$ 
\begin{eqnarray}
 P^{[2]living}_T[ \rho(.)]   \opsimeq_{T \to +\infty} 
 \delta\left( \rho(0) \right)
  \delta \left(\int_{0}^{N} dx \rho( x) -1  \right) 
 e^{- \displaystyle T I^{living}_{2} [ \rho(.)] }
\label{level2diffb}
\end{eqnarray}
where the constitutive constraints impose the normalization of the empirical density $\rho(x)$ on $[0,N]$
and its vanishing $\rho(0) =0 $ at the origin $x=0$,
while the rate function obtained from Eq. \ref{rate2.5diff} 
\begin{eqnarray}
I_{2}^{living} [ \rho(.)]  
= \int_0^{N} \frac{d x}{ 4 D( x) \rho( x) } \left[  \rho( x)  v( x)- D( x)  \rho'( x) \right]^2 
= \int_0^{N} dx \frac{\rho( x) D( x)}{ 4   } \left[  \frac{ v( x)}{D( x)} -    \frac{ \rho'( x) }{\rho(x) } \right]^2 
\label{rate2diff}
\end{eqnarray}
represents the extinction rate of a metastable state as a function of its empirical density $\rho(x)$.
Eq. \ref{rate2diff} is the analog of Eq. \ref{rate2bd} concerning the discrete birth-death model.



\end{document}